\documentclass[acmsmall,screen]{acmart}

\AtBeginDocument{%
  \providecommand\BibTeX{{%
    \normalfont B\kern-0.5em{\scshape i\kern-0.25em b}\kern-0.8em\TeX}}}

\usepackage{booktabs}   
\usepackage{subcaption} 

\usepackage{mdframed}
\usepackage{listings}

\mdfsetup{skipbelow=4pt,skipabove=4pt,leftmargin=6pt,rightmargin=0pt,align=left,usetwoside=false}

\mdfdefinestyle{listingstyle}{
  backgroundcolor=black!2,
  linewidth=2pt,linecolor=black!20,
  outerlinewidth=5pt,outerlinecolor=black,
  rightline=false,topline=false,bottomline=false,
  innerleftmargin=6pt,innerrightmargin=2pt,innertopmargin=0pt,innerbottommargin=0pt,
}

\surroundwithmdframed[style=listingstyle]{lstlisting}

\makeatletter
\lst@AddToHook{OnEmptyLine}{\vspace{\dimexpr-\baselineskip+2\smallskipamount}}
\makeatother

\usepackage{algorithm}
\usepackage[noend]{algpseudocode}
\usepackage{algorithmicx}
\algblockdefx{RepeatUntilTimeout}{EndRepeat}{\textbf{repeat until timeout}}{}
\algblockdefx{RepeatUntil}{EndRepeat}{\textbf{repeat until }}{}
\algtext*{EndRepeat}
\usepackage{float}
\newfloat{algorithm}{t}{}

\algnewcommand\algorithmicinput{\textbf{Input:}}
\algnewcommand\Input{\item[\algorithmicinput]}
\algnewcommand\algorithmicoutput{\textbf{Output:}}
\algnewcommand\Output{\item[\algorithmicoutput]}
\algnewcommand\algorithmicdata{\textbf{Auxiliary Data:}}
\algnewcommand\Data{\item[\algorithmicdata]}
\makeatletter
\algnewcommand{\LineComment}[1]{\Statex \hskip\ALG@thistlm \(\triangleright\) #1}
\makeatother

\usepackage{etoolbox}
\usepackage{tikz}
\usetikzlibrary{tikzmark}
\usetikzlibrary{calc}

\newcommand{\ALGtikzmarkcolor}{lightgray}
\newcommand{\ALGtikzmarkextraindent}{4pt}
\newcommand{\ALGtikzmarkverticaloffsetstart}{-.7ex}
\newcommand{\ALGtikzmarkverticaloffsetend}{-.5ex}
\makeatletter
\newcounter{ALG@tikzmark@tempcnta}

\newcommand\ALG@tikzmark@start{%
    \global\let\ALG@tikzmark@last\ALG@tikzmark@starttext%
    \expandafter\edef\csname ALG@tikzmark@\theALG@nested\endcsname{\theALG@tikzmark@tempcnta}%
    \tikzmark{ALG@tikzmark@start@\csname ALG@tikzmark@\theALG@nested\endcsname}%
    \addtocounter{ALG@tikzmark@tempcnta}{1}%
}

\def\ALG@tikzmark@starttext{start}
\newcommand\ALG@tikzmark@end{%
    \ifx\ALG@tikzmark@last\ALG@tikzmark@starttext
    \else
        \tikzmark{ALG@tikzmark@end@\csname ALG@tikzmark@\theALG@nested\endcsname}%
        \tikz[overlay,remember picture] \draw[\ALGtikzmarkcolor] let \p{S}=($(pic cs:ALG@tikzmark@start@\csname ALG@tikzmark@\theALG@nested\endcsname)+(\ALGtikzmarkextraindent,\ALGtikzmarkverticaloffsetstart)$), \p{E}=($(pic cs:ALG@tikzmark@end@\csname ALG@tikzmark@\theALG@nested\endcsname)+(\ALGtikzmarkextraindent,\ALGtikzmarkverticaloffsetend)$) in (\x{S},\y{S})--(\x{S},\y{E});%
    \fi
    \gdef\ALG@tikzmark@last{end}%
}

\apptocmd{\ALG@beginblock}{\ALG@tikzmark@start}{}{\errmessage{failed to patch}}
\pretocmd{\ALG@endblock}{\ALG@tikzmark@end}{}{\errmessage{failed to patch}}
\makeatother

\algnewcommand{\LeftComment}[1]{\Statex \(\triangleright\) #1}

\algrenewcommand\algorithmicindent{1em}

\usepackage{bm}

\newcommand{\tool}{TF-Coder}

\newcommand{\code}[1]{{\small \texttt{\color{blue!40!black}{#1}}}}

\setcopyright{rightsretained}

\begin{document}

\acmJournal{TOPLAS}
\acmYear{2022} \acmVolume{44} \acmNumber{2} \acmArticle{10} \acmMonth{3} \acmPrice{}\acmDOI{10.1145/3517034}

\title{\tool: Program Synthesis for Tensor Manipulations}

\author{Kensen Shi}
\affiliation{
  \institution{Google Brain}
  \country{United States}
}
\email{kshi@google.com}

\author{David Bieber}
\affiliation{
  \institution{Google Brain}
  \country{United States}
}
\email{dbieber@google.com}

\author{Rishabh Singh}
\affiliation{
  \institution{Google Brain}
  \country{United States}
}
\email{rising@google.com}

\begin{abstract}
The success and popularity of deep learning is on the rise, partially due to powerful deep learning frameworks such as TensorFlow and PyTorch that make it easier to develop deep learning models. However, these libraries also come with steep learning curves, since programming in these frameworks is quite different from traditional imperative programming with explicit loops and conditionals. In this work, we present a tool called \tool\ for programming by example in TensorFlow. \tool\ uses a bottom-up weighted enumerative search, with value-based pruning of equivalent expressions and flexible type- and value-based filtering to ensure that expressions adhere to various requirements imposed by the TensorFlow library. We train models to predict TensorFlow operations from features of the input and output tensors and natural language descriptions of tasks, to prioritize relevant operations during search. \tool\ solves 63 of 70 real-world tasks within 5 minutes, sometimes finding simpler solutions in less time compared to experienced human programmers.
\end{abstract}



\maketitle

\section{Introduction}

Deep learning techniques have resulted in recent breakthroughs in many domains including computer vision, audio processing, natural language processing, and robotics~\cite{deeplearning}. These breakthroughs arise through a combination of advancements including new algorithmic ideas, the availability of large labeled datasets, and specialized hardware for efficient training. Playing an equally important role are deep learning frameworks such as TensorFlow~\cite{tensorflow}, PyTorch~\cite{pytorch}, MXNet~\cite{mxnet}, and CNTK~\cite{cntk} that enable machine learning researchers and engineers to develop and iterate on such models more effectively.

While these deep learning frameworks have greatly eased the development and training of complex neural network models, they also have a steep learning curve, since the programming paradigm of computing over tensors using a fixed set of library functions is quite different from the traditional imperative programming paradigm. For instance, vectorization techniques are used to turn explicit loops into more efficient tensor operations, and special operations like \code{tf.where} are used in place of traditional \code{if}/\code{else} conditionals. Most deep learning models require various \emph{tensor manipulations} for data processing or cleaning, custom loss functions, and accuracy metrics, that all must be implemented within the constraints of the chosen deep learning framework. Furthermore, these frameworks offer a huge amount of functionality, which makes them powerful but potentially difficult to navigate. For instance, there are nearly 2000 distinct symbols in TensorFlow (including aliases), and about 500 of them are tensor-manipulating operations, so finding the right ones to use for a given task can be a challenge itself.

Given the increasing popularity of deep learning, combined with the relative difficulty of writing neural models, many beginners and even experienced software engineers seek assistance from others by asking questions on forums like StackOverflow. Tensor manipulations are a common difficulty, and such questions typically include an \emph{input/output example} illustrating the desired computation or transformation along with a \emph{natural language description} of what the asker is trying to accomplish. This is usually enough information for a generous expert to answer the question by providing code that implements the desired functionality, but not all questions are lucky enough to receive a correct answer or even an answer at all.

Inspired by this need, we present \tool, a programming by example system to automatically synthesize tensor manipulation programs from input/output examples and (optionally) natural language descriptions. The synthesis algorithm in \tool\ builds upon the bottom-up enumerative algorithm proposed earlier in \textsc{Transit}~\cite{transit}. In particular, we introduce three key ideas in the synthesis algorithm. First, we introduce per-operation weights to the prior algorithm, allowing \tool\ to enumerate over TensorFlow expressions in order of increasing complexity (instead of only the expression size). Second,
we introduce a novel, flexible, and efficient \emph{type- and value-based filtering system} that handles arbitrary constraints imposed by the TensorFlow library, such as ``the two tensor arguments must have broadcastable shapes.'' Finally, we develop a framework to combine predictions from multiple independent machine learning models that choose operations to prioritize during the search, conditioned on features of the input and output tensors and a natural language description of the task. This helps tailor the search process to fit the particular synthesis task at hand, as well as enables \tool\ to support multi-modal specifications (input/output examples and natural language descriptions).

The domain of tensor manipulations has not been considered previously in the program synthesis literature. This domain is particularly challenging as it encompasses a huge variety of tasks, including reshapes, filters, aggregations, maps, indexing, slicing, grouping, sorting, mathematical operations, and combinations of them. When mathematical operations (e.g., tensor products) are involved, the output tensor typically has no overlapping entries with the input tensors, ruling out synthesis approaches that are informed by partial matches between the inputs and outputs, as is common in manipulation of tables~\cite{autopandas}, data structures~\cite{lambdasquare}, and strings~\cite{flashfill}.

It is important to consider the size of the program space for this domain. Out of several hundreds of TensorFlow operations, our work supports 134 different operations, most of which take multiple arguments. Tensor manipulation problems typically involve multiple input tensors and a variety of constants as well, such as axis dimensions, tensor shapes, and tensor data types like \code{tf.int32}. Furthermore, most operations impose detailed constraints on their arguments (which cannot be formalized in SMT theories, or would require expensive non-linear reasoning). Difficult problems in this domain involve a composition of four or five different operations and expression trees of 10 or more nodes, with little room for error as shapes and data types must be compatible throughout. What synthesis techniques might we expect to handle this enormous program space?

A major contribution of this work is demonstrating that a highly optimized enumerative search strategy \emph{does} scale to solve real-world tensor manipulation problems within seconds, despite the incredible difficulty of the domain. In fact, we show that \tool\ finds solutions to problems from StackOverflow that human programmers struggle to solve. Even to us, the authors of this work, the success of enumerative search is unexpected and impressive considering its conceptual simplicity and exponential runtime. To dig deeper into this discovery, we discuss why the techniques in \tool\ are particularly effective for synthesizing programs in the domain of tensor manipulation.

We evaluate \tool\ on 70 real-world tensor transformation tasks from StackOverflow and from an industrial setting. \tool\ can successfully synthesize solutions to 63 tasks in 17 seconds on average, while \textsc{Transit} only solves 39 tasks. Moreover, our framework for incorporating multiple trained models leads to significantly faster synthesis times (1.78$\times$ average speedup) compared to not using the models. We also observed that \tool\ sometimes produces solutions that are simpler and more elegant than those written by experienced TensorFlow users (including the authors of this paper).

This paper makes the following key contributions:
\begin{enumerate}
\item We introduce \tool, the first programming by example system for synthesizing tensor manipulations in TensorFlow from input/output examples.
\item We present a weighted enumerative search algorithm that uses a novel two-stage filtering approach to efficiently enforce arbitrary preconditions required by the operations.
\item We demonstrate a framework in which multiple independent machine learning models, trained to predict useful TensorFlow operations for a given problem, are combined to guide the weighted enumerative search.
\item Most importantly, we show that \tool\ outperforms prior synthesis techniques and solves tricky real-world tasks that experienced TensorFlow programmers struggle to solve, despite using an enumerative search in an exponential space.
\end{enumerate}

\begin{figure*}[t!]
    \centering
    \begin{subfigure}[t]{0.4\textwidth}
        \centering
        \includegraphics[scale=0.31, trim={0 4cm 0 0},clip]{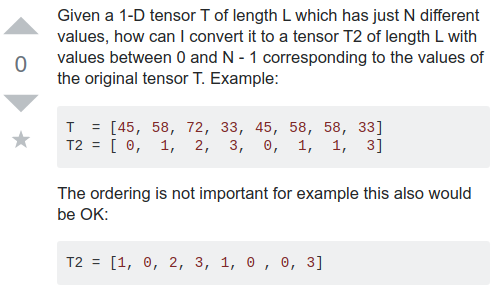}
        \caption{Labeling duplicate values in a tensor.}
        \label{fig:so_example1}
    \end{subfigure}%
    \quad
    \begin{subfigure}[t]{0.57\textwidth}
        \centering
        \includegraphics[scale=0.32]{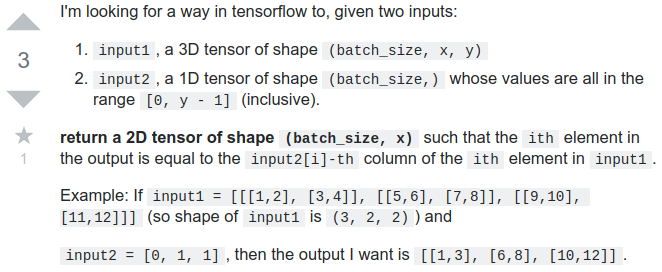}
        \caption{Gather columns of a tensor using indices in another tensor.}
        \label{fig:so_example2}
    \end{subfigure}
    \vspace*{-3mm}
    \caption{Two example tensor transformation tasks in StackOverflow posts.\protect\footnotemark}
\end{figure*}

\section{Motivating Examples}
\label{sec:motivating}

We now present some tensor manipulation questions posted to StackOverflow, an online programming help forum.

\footnotetext{
\url{https://stackoverflow.com/q/53054668} and \url{https://stackoverflow.com/q/54274074},
used under the Apache 2.0 License.}

\subsection{Example 1}
\label{subsec:example_1}

Consider the StackOverflow question shown in Figure~\ref{fig:so_example1}. The user has a 1-dimensional tensor of length $L$ containing $N \le L$ distinct values, and they want to create another tensor of the same shape containing values between $0$ and $N-1$, such that both tensors have duplicate values at the same locations. The user provides a clarifying example: the tensor \mbox{\code{[45, 58, 72, 33, 45, 58, 58, 33]}} should be converted to \mbox{\code{[0, 1, 2, 3, 0, 1, 1, 3]}}. For this problem, \tool\ synthesizes a solution program in 0.9 seconds:

\begin{lstlisting}
output = tf.unique_with_counts(in1)[1]
\end{lstlisting}

Even though the solution is relatively simple, it would be quite difficult for the question asker to find that solution without assistance, considering that there are about 500 tensor-manipulating operations in TensorFlow. Even searching for the function by name would be difficult, as the name ``\code{unique\_with\_counts}'' bears little resemblance to the user's description of the task. In such scenarios, \tool\ can help users find relevant TensorFlow operations automatically, reducing the time spent digging through documentation.

When we first came across this question on StackOverflow, it was four days old with zero answers. We posted \tool's solution as an answer, which was accepted by the poster.

\subsection{Example 2}
\label{subsec:example_2}

The StackOverflow question in Figure~\ref{fig:so_example2} involves a more difficult problem. Given two input tensors \code{in1} and \code{in2}, the question asker wants an output tensor where the \code{i}${}^\text{th}$ element is equal to the \code{in2[i]}${}^\text{th}$ column of \code{in1[i]}. To specify their intent more clearly, the asker also provides an input/output example as shown in the figure. On this complex problem involving multiple input tensors and three different TensorFlow operations, \tool\ finds a solution in 54 seconds:
\begin{lstlisting}
output = tf.squeeze(tf.gather(in1, tf.expand_dims(in2, 1), axis=-1, batch_dims=1))
\end{lstlisting}
\tool's solution is actually simpler than the accepted StackOverflow answer. Thus, \tool\ can help users find elegant solutions for difficult tensor transformations.

\subsection{Observations}

These StackOverflow questions follow a larger pattern: many tensor transformations are ambiguous if described using natural language alone, so it is natural to provide both a textual description of the desired transformation and concrete input/output example tensors to clarify the problem. Another interesting property is that most of the time, only one input/output example is necessary, since tensors can be expanded with more entries to resolve ambiguities.

There are over 60,000 questions on StackOverflow containing the text ``TensorFlow.'' While the majority of these ask about installation issues or deep learning in general, there are still many questions asking how to perform tensor manipulations or how to fix errors raised by the user's code. Indeed, writing TensorFlow code can be challenging at times (even more so for beginners) due to the amount of information that the programmer must keep in mind. The shapes of tensors must be checked for compatibility under broadcasting rules, the conceptual meanings of the dimensions are crucial to ensure mathematical correctness, and data types of tensors must be tracked carefully (e.g., a \code{tf.int32} tensor cannot be added to a \code{tf.int64} tensor). Furthermore, these properties change as tensors are manipulated, leaving many opportunities for subtle bugs.

Inspired by these questions, we developed \tool\ to automatically synthesize tensor manipulations in TensorFlow from input/output examples and (optionally) natural language descriptions. Such a tool could help accelerate users' TensorFlow development in several ways. In Section~\ref{subsec:example_1}, we observed that \tool\ can automatically find relevant TensorFlow operations, thus reducing the need to search through TensorFlow's extensive documentation. Since \tool's solutions are guaranteed to be consistent with the provided input/output example, it can reduce the number of debugging cycles and lead to increased confidence in the code (much like a unit test). Finally, by finding simple and elegant solutions that the user may have overlooked, \tool\ can even improve code quality and model efficiency. We strive to find solutions quickly, within seconds or at most a few minutes, so that the tool may be used interactively.

\section{Synthesis with Enumerative Search}
\label{sec:search}

\begin{figure}[!t]
    \includegraphics[scale=0.24]{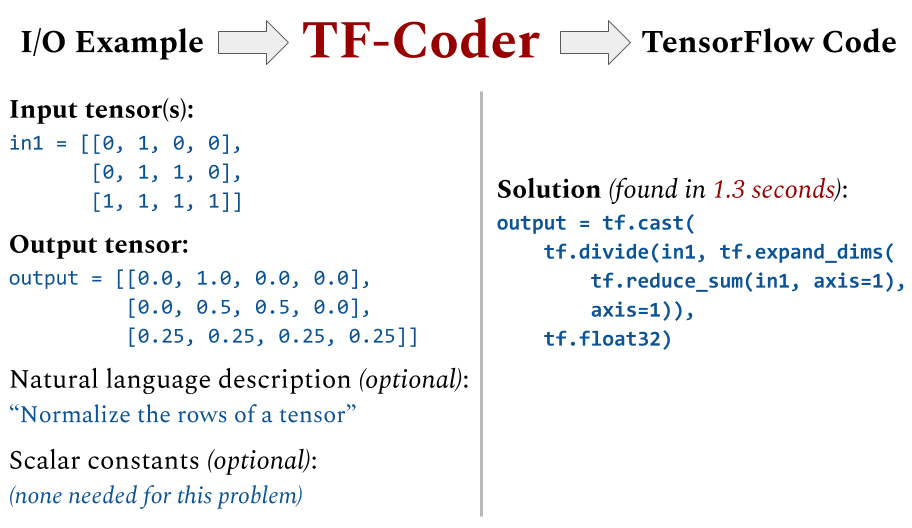}
    \vskip -8pt
    \caption{Given an input/output example of a tensor manipulation, an optional natural language description, and optional scalar constants, \tool\ synthesizes a composition of TensorFlow operations consistent with the example.}
    \label{fig:problem}
\end{figure}

Motivated by the examples and discussion in Section~\ref{sec:motivating}, we now formalize the problem as illustrated in Figure~\ref{fig:problem}.

\subsection{Problem Formalization}

We assume a given task specification $\phi = \{(\mathcal{I}, \mathcal{O}), D, C\}$, where $(\mathcal{I}, \mathcal{O})$ is an input/output example, i.e., a list of input tensors $\mathcal{I}$ and the corresponding output tensor $\mathcal{O}$, $D$ is an optional natural language description of the task, and $C$ is an optional set of constants that may be useful for the task.

Our goal is to synthesize a program $P \in \mathcal{D}$ where $P(\mathcal{I}) = \mathcal{O}$. We note that \tool\ can often synthesize programs directly from the input/output example $(\mathcal{I}, \mathcal{O})$ without \mbox{needing} additional $D$ and $C$ information, but $D$ and $C$ allow users to express their intent and obtain better synthesizer performance. The domain of programs $\mathcal{D}$ considered by \tool\ consists of single-line TensorFlow expressions, which may contain any of the following \emph{base values}:
\begin{itemize}
    \item Python int, float, Boolean, and string literals
    \item TensorFlow data types, e.g., \code{tf.float32}, \code{tf.int64} etc.
    \item Variables \code{in1}, \code{in2}, etc., to reference the input tensors
\end{itemize}
Furthermore, expressions may use the following \emph{operations}, applied to the base values or composed with each other:
\begin{itemize}
    \item Supported TensorFlow function calls, e.g., \code{tf.add(x, y)} and \code{tf.math.segment\_max(data, segment\_ids)}
    \item Creating a tuple from supported Python literals, e.g., \code{(0, 1)}, or from other such tuples
    \item Various forms of indexing and slicing of sequences and tensors, e.g., \code{tensor[-1]}, \code{tensor[1:]}, and \code{tensor[:, 0:5]}
\end{itemize}

Note that the TensorFlow operations specify their arguments because the search algorithm requires a fixed arity for each operation. Hence, some TensorFlow functions have multiple supported variations, e.g., 2-argument \code{tf.gather(params, indices)} and 4-argument \code{tf.gather(params, indices, axis, batch\_dims)}. In total, \tool\ currently supports 123 TensorFlow operations for 99 distinct functions, plus 11 more operations for different forms of indexing, slicing, and tuple creation. These are listed in Appendix~\ref{app:ops}.

In the following sections, we describe the weighted bottom-up enumerative search that powers \tool. Starting with a set of initial values including input tensors and constants (which may be provided by the user or chosen heuristically), the search enumerates ways of applying operations to previously-explored values, to expand the set of known values. Values internally store enough information to recursively reconstruct the code expression that would produce the value. Thus, if the search encounters a value that matches the output tensor, the matching value's code expression is a valid solution to the synthesis problem.

\subsection{Weighted Value Search}
\label{subsec:weighted}

\begin{figure}[!t]
    \includegraphics[scale=0.17]{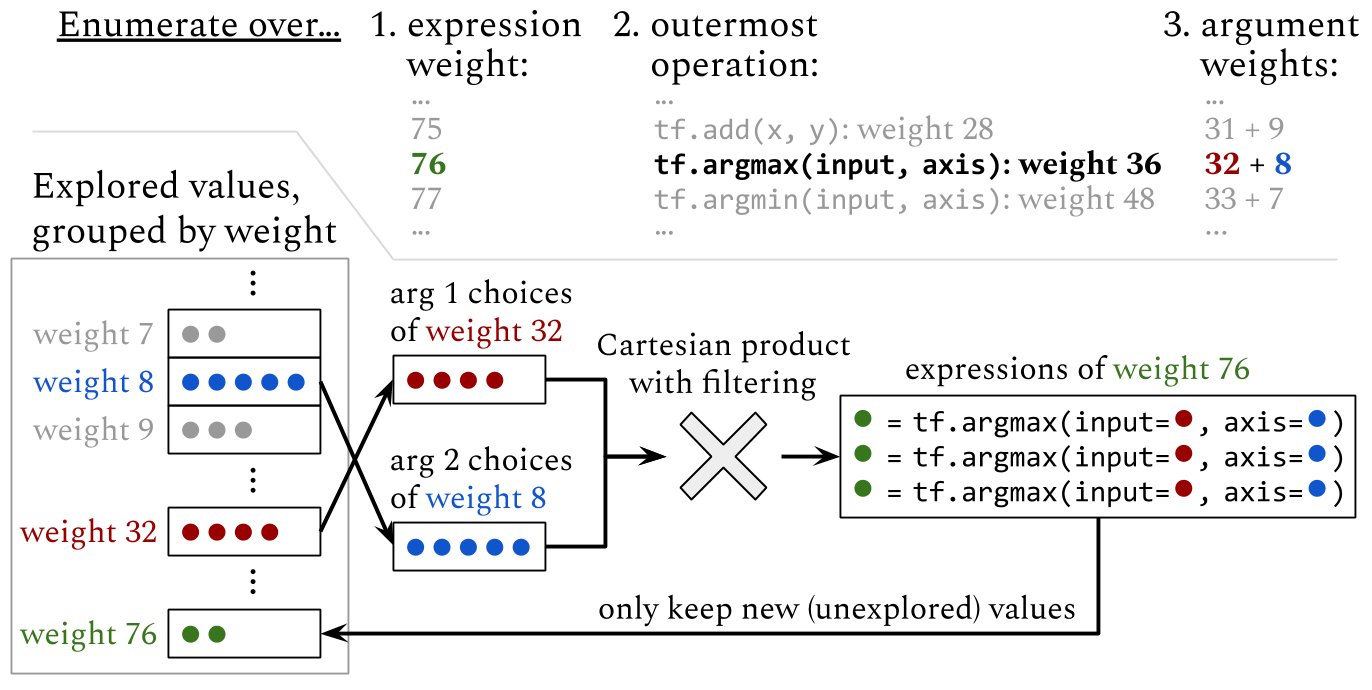}
    \vspace*{-2mm}
    \caption{\emph{Overview of the enumerative search algorithm.} \quad \tool\ stores already-explored values organized by weight, initially just the input tensors and constants.
    It enumerates expressions in order of increasing weight. For a target expression weight (e.g., 76), it enumerates over operations
    and weights for the operation's arguments, e.g., the operation \code{tf.argmax(input, axis)} has weight 36 and two arguments, so the remaining weight ($76 - 36 = 40$) is partitioned into two parts (e.g., $32 + 8$) representing the arguments' weights. Options for the arguments are drawn from previously-explored values, and a Cartesian product with customizable filtering produces lists of arguments. Finally, invoking the operation produces new values.}
    \label{fig:overview}
\end{figure}

\tool's search enumerates expressions in order of increasing \emph{weight}, which represents the expression's complexity. Operations and initial values (input tensors and constants) have associated weights, and an expression's weight is defined to be the sum of the weights of the operations and initial values used in that expression. For example, the initial values \code{in1} and \code{0} both have weight 8, and the operation \code{tf.expand\_dims(input, axis)} has weight 18, so the expression \code{tf.expand\_dims(in1, axis=0)} has weight $8 + 8 + 18 = 34$.

These weights give \tool\ a fine-grained notion of the ``complexity'' of different TensorFlow operations. For instance, \code{tf.reverse(tensor, axis)} is more complex and less useful than \code{tf.expand\_dims(input, axis)}, so the former is assigned a greater weight than the latter. We manually assigned weights for each of \tool's supported operations, taking into consideration how common or useful the operation is, how complex its semantics are, and how many arguments it takes. All weights must be positive integers to enable efficient enumeration. These weights allow \tool\ to prioritize simple and useful operations in its search, and this is crucial for enabling \tool\ to handle so many different operations---niche operations are given higher weight so they can still be used if necessary, without causing much slowdown if they are not needed in a particular problem.

Figure~\ref{fig:overview} is a diagram summarizing \tool's weighted enumerative search, and the algorithm is shown in Algorithm~\ref{alg:search}. Note that the algorithm mentions using learned models to prioritize operations, discussed in Section~\ref{sec:learning}. Argument filters and combination filters are discussed in Section~\ref{subsec:filtering}.

The algorithm starts by collecting initial values. These include user-provided input tensors, user-provided constants (optional), and heuristically-chosen constants. The constants \code{0}, \code{1}, \code{-1}, \code{True}, \code{False}, \code{tf.int32}, \code{tf.int64}, \code{tf.float32}, and \code{tf.bool} are always chosen. We also include natural numbers up to the maximum rank of an input tensor (exclusive) to serve as axis values, all dimension lengths of input and output tensors, and the output tensor's shape as a tuple. These initial values are assigned hardcoded weights depending on their origin (e.g., a user-provided constant will have smaller weight than a constant extracted from a dimension length).

The search then generates expressions in order of increasing weight. For a given target weight, it enumerates over all supported operations and all allowable weights for the operation's arguments. For example, if we are currently generating expressions of weight 76 using a 2-argument operation with weight 36, then there is $76 - 36 = 40$ remaining weight to partition among the two arguments. If argument 1 is chosen to have weight 32 and argument 2 is chosen to have weight 8, we would use all previously-explored values of weight 32 as choices to fill argument 1, and similarly all existing values of weight 8 are choices for argument 2. The Cartesian product of these argument choices gives many argument lists, each list containing one concrete value for each argument. The chosen operation is applied to each of these argument lists to produce new values, which by construction all have the desired weight. Each newly generated value that is not equal to a previously-seen value is added back to the set of known explored values. In this way, we prune away expressions with equivalent behavior when run on the input tensors to reduce the size of the search space.
Every value produced by applying an operation to arguments stores references to the operation and the arguments, so that any value can recursively reconstruct its code representation. As soon as \tool\ encounters a value that is equal to the desired output tensor, it outputs the value's code representation as a solution.

In general, if there are multiple code solutions to a programming problem, the ideal solution is both intuitive and short. If these two desires are at odds, one must find a reasonable balance between them. This is a key motivation behind our weighted search, where the total weight of a solution is reduced if it uses operations with smaller weight (the simple, intuitive, or common operations) or if it uses less code (fewer nodes in the expression tree). Hence, the weight of a solution can be used as a metric of how desirable it is, balancing the two goals of being concise and understandable. Under this formulation, \tool\ is guaranteed to find the ``best'' solution, i.e., the solution with smallest weight, because it considers expressions in order of increasing weight.

\begin{algorithm}
    \caption{\quad\tool's Synthesis Algorithm}
    \label{alg:search}
    \begin{algorithmic}[1]
        \Input Input/output example $(\mathcal{I}, \mathcal{O})$, natural language description $D$, user-provided constants $C$
        \Output A program $P$ such that $P(\mathcal{I}) = \mathcal{O}$
        \Data Supported operations \emph{Ops} (each $op$ has argument filters $\{op.f_i\}$ and combination filter $op.F$), a model $M_\mathit{io}$ conditioned on input/output examples, and model $M_\mathit{nl}$ conditioned on natural language
        \Statex \vskip -0.9em
        \LeftComment{Use learned models to prioritize operations}
        \State $m_\mathit{io} \gets M_\mathit{io}(\mathcal{I}, \mathcal{O})$ \Comment{Model predictions}
        \State $m_\mathit{nl} \gets M_\mathit{nl}(D)$
        \ForAll{$op \in \mathit{Ops}$} \Comment{Prioritize ops using models}
            \State $op.\mathit{weight} \gets $ \Call{ReweightOp}{$op, m_\mathit{io}, m_\mathit{nl}$}
        \EndFor

        \Statex \vskip -1em
        \LeftComment{Gather initial values with weights}
        \State $E \gets \mathcal{I} \cup C$ \Comment{Set of explored values}
        \State $E \gets E \cup {}$\Call{HeuristicConstants}{$\mathcal{I}, \mathcal{O}$}
        \ForAll{$v\in E$}
            \State $v.\mathit{weight} \gets $ \Call{AssignWeightByOrigin}{$v$}
        \EndFor

        \Statex \vskip -1em
        \LeftComment{Bottom-up enumerative search}
        \For{$W = 1, 2, \dots$} \Comment{Weight of expressions}
            \ForAll{$op \in \mathit{Ops}$}
                \State $n \gets op.\mathit{arity}$
                \State $w \gets op.\mathit{weight}$
                \ForAll{$[w_1, \dots, w_n]$ s.t. $\sum_i w_i = W-w, \enskip w_i \in \mathbb{Z}^+$}\Comment{Argument weights}
                    \For{$i = 1, \dots, n$} \Comment{Collect argument choices}
                        \State $A_i \gets \{e \in E \mid e.\mathit{weight} = w_i \wedge op.f_i(e)\}$
                    \EndFor
                    \ForAll{$[a_1, \dots, a_n] \in \Pi_i A_i$}
                        \If{$\neg op.F([a_1, \dots, a_n])$}
                            \State \textbf{continue}
                        \EndIf
                        \State $V \gets $ \Call{Execute}{$op, [a_1, \dots, a_n]$}
                        \If{$V \not \in E$} \Comment{New value discovered}
                            \State $V.\mathit{weight} \gets W$
                            \State $V.\mathit{history} \gets (op, [a_1, \dots, a_n])$
                            \State $E \gets E \cup \{V\}$
                        \EndIf
                        \If{$V = \mathcal{O}$} \Comment{Solution found}
                            \State \textbf{return} \Call{CodeExpression}{$V$}
                        \EndIf
                    \EndFor
                \EndFor
            \EndFor
        \EndFor
    \end{algorithmic}
\end{algorithm}

\subsection{Operation Filtering}
\label{subsec:filtering}

\begin{figure*}[!tpb]
    \includegraphics[scale=0.155]{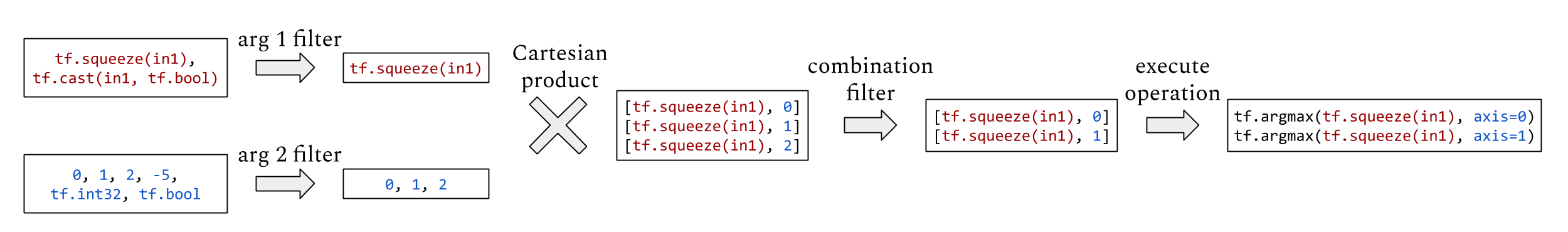}
    \caption{\emph{Two-stage operation filtering.} \quad Here we demonstrate \tool's flexible two-stage operation filtering for the \code{tf.argmax(input, axis)} operation. The first argument, \code{input}, must be a numeric tensor (e.g., not a boolean tensor), so \code{tf.cast(in1, tf.bool)} is removed by the ``arg 1 filter.'' The second argument, \code{axis}, must be an integer that is a potential axis value, so \code{-5}, \code{tf.int32}, and \code{tf.bool} are removed by the ``arg 2 filter.'' Finally, the \code{axis} must be in range for the particular \code{input}, so \code{[tf.squeeze(in1), 2]} is removed by the ``combination filter'' if \code{tf.squeeze(in1)} actually has rank 2. After these filtering steps, the \code{tf.argmax} operation is applied to the surviving combinations of arguments.}
    \label{fig:filtering}
\end{figure*}

When the search enumerates argument lists for a particular operation, a full Cartesian product of argument choices may be very large, even though very few argument lists actually meet preconditions required by the operation. To avoid enormous Cartesian products, and to reduce the number of errors thrown by operations (which are relatively expensive to catch), we introduce a flexible two-stage operation filtering approach, illustrated in Figure~\ref{fig:filtering}.

The first stage of operation filtering occurs independently for each argument of the operation. An ``argument filter'' ($op.f_i$ in Algorithm~\ref{alg:search}) is simply a function that takes a value and returns a boolean denoting whether the value is an acceptable choice for a particular argument of an operation. For example, the \code{tf.argmax(input, axis)} operation requires that the \code{input} argument be a numeric tensor (e.g., a tensor with a float or int data type), and the \code{axis} argument must be an integer representing an axis. Hence, an argument filter for \code{input} would reject tensors with \code{tf.bool} data types, and an argument filter for \code{axis} would only accept integers with small absolute value. By using argument filters, the size of the Cartesian product of argument choices is greatly reduced.

The second stage of operation filtering checks constraints that involve multiple arguments. A ``combination filter'' ($op.F$ in Algorithm~\ref{alg:search}) for an operation with $n$ arguments is a function that takes a list of $n$ values and returns a boolean denoting whether the list contains acceptable arguments for one call to the operation. For example, the \code{tf.argmax(input, axis)} operation requires that the axis actually be in range for the input tensor. Hence, the operation's combination filter would remove an argument list if it has an out-of-bounds axis for the corresponding input tensor. The purpose of combination filters is to avoid executing expensive TensorFlow operations that can be eliminated by quick checks. Furthermore, catching exceptions raised by TensorFlow operations is relatively slow compared to running the combination filter.

The two-stage filtering approach allows for arbitrary value-based checking of operation preconditions. \tool\ is also engineered such that it is easy to add and reuse filters with minimal code duplication---many operations have an \code{axis} argument that requires the same argument filter, and similar operations like \code{tf.reduce\_sum(input\_tensor, axis)} can use the same argument and combination filters.

Finally, we note that argument filters (but not combination filters) will be run repeatedly on the same values for two reasons. First, argument filters like the \code{axis} argument filter are reused among several operations. Second, the same argument will be assigned values of the same weight at different points in the enumerative search. Our solution is to cache the result of applying an argument filter on all explored values of a given weight, i.e., we cache $A_i$ in Algorithm~\ref{sec:search}, where the cache is keyed by the filter function $op.f_i$ and the weight $w_i$ of the values being filtered. (For simplicity, this caching behavior is not present in Algorithm~\ref{alg:search}.)

\tool's operation filtering significantly improves the quality of candidate programs considered. In particular, for the difficult task described in Figure~\ref{subfig:stackoverflow_48}, overall the argument filters eliminated 73\% of choices for individual arguments, and then the combination filters further eliminated 60\% from the Cartesian product of remaining argument choices. Together, the two-stage filtering strategy eliminates 98.6\% of all potential candidate programs for a \emph{single} operation, and a correct program must satisfy the constraints for \emph{all} operations in the expression. This is evidence that our filtering strategy is particularly useful for the domain of tensor manipulations, where only a tiny fraction of candidate programs actually meet all of the required constraints.

\subsection{Domain-Specific Details}

Here we describe a few techniques in \tool\ that take advantage of the TensorFlow domain, and that may or may not be useful in other similar domains. These techniques are excluded from Algorithm~\ref{alg:search} for simplicity.

We impose limits on the sizes of values (e.g., number of elements in a tensor) encountered during search. This is done to avoid excessive memory usage through the creation of huge tensors. These limits are enforced during operation filtering, e.g., do not call \code{tf.ones(shape)} on the argument \code{tf.range(1, 20)}, as that would cause an out-of-memory error. The limits are also checked after new values are created as a blanket safeguard against memory issues, and values that are too large are immediately discarded. In our experiments, we allow tensors to have a maximum of 1000 elements, 4 dimensions, and 100 elements along a single dimension. These limits are chosen to admit the largest tensors that we expect average users to require.

Many tasks require a \code{tf.cast} operation as the final step. Instead of waiting for the \code{tf.cast} operation to be applied through the search, \tool\ opportunistically casts newly generated values to the target output's data type if the new value matches the output's shape but not its data type. If the casted value does not match the output, it is discarded. This step takes negligible time since it is applied to few values, but it drastically reduces the synthesis time for tasks that require a \code{tf.cast} as the final operation. Note that the \code{tf.cast} operation is still treated normally within the value search, which is necessary to produce and store casted values to be used as arguments to other operations later in the search.

A \emph{SparseTensor} is a special kind of tensor object that represents large sparse tensors in a memory-efficient way. TensorFlow's \code{tf.sparse} submodule is dedicated to manipulating SparseTensors, e.g., the \code{tf.add} function does not support adding SparseTensors, and the \code{tf.sparse.add} function must be used instead. Because sparse operations may be confusing to users who are not familiar with SparseTensors, we prevent all \code{tf.sparse.*} operations from being used unless a SparseTensor is given as an input or output tensor, or the description includes the term ``sparse''. This also reduces the search space for tasks that do not use SparseTensors.

\paragraph{Handling Multiple I/O Examples} In the tensor manipulation domain, we observe that most tasks only require a single input/output example. For instance, when performing a reduction across rows of an $M \times N$ matrix to produce a length-$M$ vector, there are essentially $M$ independent examples of a row being reduced to a scalar. One can easily construct a single example with large enough $M$ to unambiguously specify the task. This idea generalizes to nearly all tensor manipulation tasks: adding more numbers to the example makes it more clear.

Even so, \tool's enumerative search algorithm can be extended to handle multiple examples. To do this, we simply need to extend the notion of a ``value'' in our value search. In the single-example case, a ``value'' represents one code expression and contains the result of running that code expression using the example's inputs. In the multi-example case, a ``super-value'' still represents one code expression, but it contains the results of running that code expression on inputs from each example. For equivalence-based pruning (line 20 of Algorithm~\ref{alg:search}), two super-values are considered equal if all pairs of contained results are equal. For operation filtering (lines 15 and 17), a super-value is permitted by a filter if all of its contained results pass the filter. A solution is found (line 24) when the super-value's contained results all match the examples' outputs.

\section{Learning to Guide the Search}
\label{sec:learning}

In Section~\ref{subsec:weighted}, we noted that operation weights allow \tool\ to prioritize simple and useful operations. Another benefit is that weights can be modified to fit the specific synthesis problem at hand, instead of having static weights that are independent of the problem. This enables strategies that tweak the ordering of the search space to better fit the problem.

We propose a simple but effective framework for guiding the search using predictions from multiple independently-trained machine learning models. \tool\ uses two such models: a neural model conditioned on features of the input and output tensors, and a bag-of-words model conditioned on the natural language description of the problem. These models draw signal from completely different aspects of the problem. Our experiments in Section~\ref{subsec:model_results} show that each of the two models alone leads to a noticeable performance improvement, which is only increased when the models are used together in our framework. Although we only explore these two models in this paper, our framework easily generalizes to incorporate more models that would provide signals from other sources, e.g., the surrounding code context where the tensor manipulation is needed.

A key property of our framework is that each model is trained and used independently. Because there currently does not exist a large dataset of TensorFlow expressions along with input/output examples and natural language descriptions, we cannot directly train a single unified model to understand input/output examples and natural language descriptions simultaneously. Instead, by using separate independent models, we can use separate training datasets that are more readily available. Furthermore, any model in the framework can be optional, so multi-modal specifications can be supported without being required. For instance, a single unified model might not handle empty descriptions well, but the predictions of a separate natural language model can simply be ignored if a description is not provided.

Our framework works in the following way. Each model's predictions are used to prioritize the chosen operations by multiplying their weights by a constant. This is $0.75$ in our experiments for simplicity, but in general each model may have a separate constant multiplier (a hyperparameter) that represents how trustworthy the model is, considering that the models are conditioned on different parts of the problem instance. The models independently choose which operations to prioritize, so if an operation is prioritized by both models, its weight will be multiplied by $0.75^2$. Modified weights are rounded to the nearest integer (or rounded up to 1 since weights must be positive). Then, the enumerative search described in Section~\ref{sec:search} is run as normal, using the modified operation weights.

\subsection{Tensor Features Model}
\label{subsec:tensor_model}

We now describe a neural model that learns a Bernoulli distribution over each operation, conditioned on features of input and output tensors. Human programmers can often recognize useful operations for tensor transformation tasks by looking at patterns in the user-provided examples. For instance, if one tensor contains small nonnegative integers, they may represent indices into another tensor, especially if the output tensor also contains entries that are found in the input tensors. Or, if the output tensor does not have many entries that also appear in some input tensor, then the desired transformation is likely to include some mathematical operation such as a tensor product. With the tensor features model, our goal is to learn such pattern-recognition capabilities.

\paragraph{Dataset}

One challenge for training such a model is the lack of a large supervised dataset containing real TensorFlow programs together with corresponding input/output examples, so we train our model on a synthetically generated dataset. However, unlike previous approaches~\cite{robustfill, deepcoder, syntheticdataset} that uniformly sample from a space of programs and inputs, we observe that this approach in the TensorFlow domain will result in a huge number of errors due to the many constraints imposed by TensorFlow operations. Furthermore, without symbolic formulas for these constraints, we cannot use solver-based approaches to find satisfactory programs and inputs~\cite{symex, klee}.

We present a novel idea of generating the synthetic training dataset using our enumerative search algorithm, running the weighted value search on randomly-generated inputs for 10 minutes to gather a large number of explored values. For each such value, we consider all ways of collapsing subtrees of its code expression into new inputs, to add more variety in the input tensors. For example, given the code expression \code{tf.greater(tf.add(in1, tf.squeeze(in2)), in3))}, we would additionally consider the expressions \code{tf.greater(new\_input, in3)} and \code{tf.greater(tf.add(in1, new\_input), in3))}, where \code{new\_input} is a new input tensor with a value equal to the value of the code subtree that it replaced. We randomly choose one such way of collapsing subtrees (including the original expression unchanged) for each explored value, resulting in an I/O example with a corresponding TensorFlow program.

We then filter the dataset to only contain programs that use at least two operations, since programs using one single operation are already easily synthesized by the value search in a fraction of a second. Additionally, we also exclude examples where an input or output tensor has more than 50 elements, to more closely resemble example tensors that would be manually provided by \tool's users. Our training dataset comes from 20,000 runs of value search on random inputs, and we randomly choose at most 2,000 explored values from each run and draw one training example from each chosen value, for a total of 39,930,863 training examples. The evaluation dataset uses 1,000 runs of value search and at most 100 examples from each run, for a total of 99,852 evaluation examples.

\paragraph{Example Features}

We compute a set of features for the input/output tensors to feed into the model, which include:
\begin{itemize}
    \item If the value is a primitive, sequence, tensor, or SparseTensor
    \item The value's data type, rank, and dimension lengths
    \item Statistics (e.g., max, min, mean) of its elements
    \item The number and fraction of elements of various properties, e.g., exactly zero, in the range $[0, 1]$, unique elements, etc.
    \item Various Boolean properties of the value, e.g., all elements positive, unique, sorted, etc.
\end{itemize}
In addition to featurizing the individual input and output tensors, we also compute features representing the comparison of each input value to the output value:
\begin{itemize}
    \item Comparing the number of elements, ranks, and each dimension length
    \item The number and fraction of input elements that also appear in the output, and vice versa
    \item If all input elements appear in output, and vice versa
    \item If each dimension length of the input also appears as some dimension length of the output, and vice versa
\end{itemize}
For all features that result in an unbounded integer or float (e.g., the maximum element or number of unique elements), we bucket the feature to turn it into a categorical feature.

To featurize an input/output example, we first pad the list of inputs with dummy input values until there are exactly 3 inputs, so that the same number of features are extracted for each example.\footnote{This scheme supports a maximum of 3 inputs, but this could be relaxed. We have not yet encountered a reasonably-complex task requiring 4 inputs.} We then extract features for each input and the output individually, and extract features from a comparison of each input to the output. We also add a single feature representing the number of inputs.

\paragraph{Models}

Our neural model first embeds categorical features (e.g., boolean properties, bucketed numbers, data types, etc.) using an embedding size equal to the cardinality of the feature. The embeddings are concatenated along with unembedded features (e.g., fraction features), resulting in a vector of length 2049. This is passed through 1 or 2 dense layers, a final dense output layer produces a logit for each operation, and elementwise sigmoid is applied to get a probability for each operation.

We experiment with different loss functions. One is a standard sigmoid cross entropy loss averaged over the operations. However, as each example only uses a few operations, the dataset is overwhelmingly negative, which could lead the model to be overly conservative with its predictions. Thus, we also implement a differentiable $F_\beta$ metric~\cite{fbeta} as a loss function to achieve different balances in precision and recall. $F_1$ prioritizes precision and recall equally, while $F_2$ prioritizes recall twice as much as precision. In general, we found that correctly prioritizing an operation outweighs prioritizing an operation that is actually not used.

There is a large imbalance in the frequency of operations appearing in ground-truth programs in the synthetic dataset. For instance, \code{tf.expand\_dims(input, axis)} appears in about 9 million ground-truth programs (out of about 40 million total tasks), while \code{tf.reshape(tensor, shape)} only appears in about 6 thousand ground-truth programs. There are several factors leading to this frequency imbalance. Some operations may have highly restrictive constraints on their arguments, such that the operation actually succeeds in a significantly smaller proportion of the time. Some operations may require complex arguments that are more difficult to construct, and so the operation begins seeing usage later during the search. For example, the constraint on \code{tf.expand\_dims}'s \code{axis} argument (it is an integer in range) is less strict than the constraints on \code{tf.reshape}'s \code{shape} argument (it is a sequence of integers representing a shape that \code{tensor} could be reshaped into). Furthermore, constructing a valid \code{axis} is very easy (usually a single integer constant) compared to constructing a valid \code{shape} (usually involving a tuple creation operation and multiple integer constants). Combined, these effects lead to \code{tf.reshape} appearing 3 orders of magnitude fewer times than \code{tf.expand\_dims}.

This frequency imbalance may be problematic if the model learns to mostly predict frequently-occurring operations like \code{tf.expand\_dims} simply because of their prevalence in the training dataset. Instead, we would like the model to base its predictions not on the frequency of operations in the dataset, but on actual patterns observed using the featurization of the I/O example. To do this, we experiment with applying a multiplicative scaling factor to the loss terms for operations used in a training example, scaling up the loss for uncommon operations and/or scaling down the loss for common operations. More precisely, when an operation $\mathit{op}_i$ appears in the current training example's ground-truth program, we multiply the training loss term corresponding to $\mathit{op}_i$ by a scaling factor of either
\[
\alpha^{\text{max}}_i = \frac{\max_j\{\text{count}(\mathit{op}_j)\}}{\text{count}(\mathit{op}_i)} \text{ or }
\alpha^{\text{mean}}_i = \frac{\text{mean}_j\{\text{count}(\mathit{op}_j)\}}{\text{count}(\mathit{op}_i)}
\]
where $\text{count}(\mathit{op}_i)$ is the number of training examples where $\mathit{op}_i$ is used in the ground-truth program, and $j$ ranges over all supported operations so that $\max_j\{\text{count}(\mathit{op}_j)\}$ and $\text{mean}_j\{\text{count}(\mathit{op}_j)\}$ are both constant properties of the entire training set. Finally, we clip $\alpha^{\text{max}}_i$ and $\alpha^{\text{mean}}_i$ to a maximum of 10,000 to avoid training instability from extremely large scaling factors. We call these two scaling approaches ``$\alpha^{\text{max}}$ scaling'' and ``$\alpha^{\text{mean}}$ scaling.''

The scaling approaches change the amount of positive training signal (i.e., that an operation should be predicted) given to the model for $\mathit{op}_i$. With $\alpha^{\text{max}}$ scaling, each time $\mathit{op}_i$ appears in a ground-truth program, the model takes $\alpha^{\text{max}}_i$ times larger of a gradient step toward predicting that $\mathit{op}_i$ is used. Over the entire training set, this approximates the scenario where $\mathit{op}_i$ appears in $\alpha^{\text{max}}_i$ times as many ground-truth programs, namely $\alpha^{\text{max}}_i \cdot \text{count}(\mathit{op}_i) = \max_j\{\text{count}(\mathit{op}_j)\}$ ground-truth programs. This is a constant for every $\mathit{op}_i$, so the operation frequency imbalance is effectively removed. Similar logic applies for $\alpha^{\text{mean}}$ scaling as well.

Using $\alpha^{\text{max}}$ scaling has the effect that no operation's loss terms are scaled down, but it leads to the model receiving more total positive training signal than in the unscaled case. In contrast, with $\alpha^{\text{mean}}$ scaling, the total amount of positive training signal obtained during a training epoch is unchanged, but there is less positive training signal for frequently-appearing operations which could be seen as wasting training data.

Considering sigmoid cross entropy, $F_1$, and $F_2$ loss functions, along with $\alpha^{\text{max}}$ scaling, $\alpha^{\text{mean}}$ scaling, or no scaling at all, we have 9 different variations. For each variation, we ran a hyperparameter sweep and selected the run with the lowest evaluation loss after 3 epochs. We observed no overfitting. We varied the number of hidden feedforward layers (1 or 2), the size of the hidden layers (512, 1024, or 2048), and the learning rate (7 choices between 1e-5 and 1e-3). We used the Adam optimizer \cite{adamopt} with global norm gradient clipping~\cite{clipping}. Results are discussed in Section~\ref{subsec:model_results}.

For all variations of the the tensor features model, we prioritize all operations where the predicted probability is greater than 0.5. We considered using the predicted probabilities to reweight prioritized operations in a more gradual way, but we observed that the trained tensor features models are very confident such that almost all predicted probabilities are close to 0 or 1. Thus, the two approaches to reweighting prioritized operations would lead to very similar outcomes.

\subsection{Natural Language Model}
\label{subsec:nl_model}

We next describe a model that prioritizes operations based on the natural language description accompanying the input/output examples. As with the tensor features model, we formulate the training objective as a supervised multilabel classification problem. Given a task's natural language description, the model predicts whether each operation occurs in the task's solution.

\paragraph{Dataset}
Since we do not have a large dataset of \tool\ queries paired with TensorFlow operations, we construct a dataset for training the natural language model from the TensorFlow documentation and from TensorFlow code on GitHub.

For each operation supported by \tool, we construct an instance for our dataset using the TensorFlow documentation. The docstring of the operation serves as the task description, and we consider the operation to be the sole target operation for the instance. This yields 134 descriptions paired with target operations.

For the remainder of the dataset, we construct examples from TensorFlow code from GitHub. We collect 65,617 functions that use at least one \tool\ supported TensorFlow operation from GitHub projects with a permissive license. Following the method of \citet{deduplication}, we remove duplicate and near-duplicate instances from this dataset, leaving 13,960 functions.
For each function, we extract a \emph{natural language context} from the function, as well as the set of supported TensorFlow operations used by the function.
The natural language context consists of the function's docstring and all comments, string literals, and variable names appearing in the function.
We use this natural language context as a proxy for the task description, and we use the TensorFlow operations found in the function as the target TensorFlow operations. In total, our full constructed dataset has 14,094 instances.

\paragraph{Models}
We consider two classes of models for the natural language model: TF-IDF cosine similarity (the TF-IDF model) and na\"ive Bayes. All models accept natural language description $D$ and operations $op_1, \ldots, op_n$ as input, and decide which operations to prioritize in the search. We restrict our models to prioritizing at most $k$ operations with the best scores. These models are implemented using scikit-learn~\cite{scikit-learn}.

Though the natural language in the constructed dataset often differs in structure from real \tool\ task descriptions, we hypothesize that we can still learn from the \emph{vocabulary} used in the dataset to perform well on the benchmark tasks. So, we focus our efforts on these two bag-of-words models, rather than training higher capacity models which would better fit the dataset but not generalize to the target domain of \tool\ task descriptions.

We train the TF-IDF model using only the TensorFlow documentation, not the instances gathered from GitHub. We construct a vocabulary consisting of those terms appearing at least once in the docstrings of the supported TensorFlow operations, with English stop words removed.
For each operation $op_i$, we construct a vector $V_{op_i}$ from the operation's docstring consisting of the tf-idf score of each term in the vocabulary \cite{tfidf}.
The tf-idf score for a term in a docstring is computed as the number of occurrences of the term in the docstring, divided by the smoothed log total number of occurrences of the term across all docstrings.
The smoothing is equivalent to there being a single extra docstring containing every term exactly once.

We construct an analogous vector $V_D$ from the input text $D$.
For natural language $D$ and operation $op_i$, the TF-IDF model produces a score given by the cosine similarity between $V_D$ and $V_{op_i}$.
The model prioritizes the operations with the highest scores, considering only those operations with score exceeding a threshold $\mathit{minScore}$, and up to $k$ operations prioritized.

We next train the na\"ive Bayes model, using the full constructed dataset. This model uses the same vocabulary and document frequencies as the TF-IDF model and the same definition of $V_D$. Though the dataset is now larger, we do not expand the vocabulary to include novel terms. We find that restricting the capacity of the model in this way reduces overfitting to the domain of the constructed dataset.

For each operation $op$, let $Y_{op}$ be a binary random variable indicating whether $op$ is used in the target program.
The na\"ive Bayes model estimates the probability of $op$ being used given natural language $D$ as
\[P(Y_{op} \mid D) \propto P(Y_{op}) \prod_{i} P(D_i \mid Y_{op}).\]
We calculate this using the estimate $P(D_i \mid Y_{op}=1) = \frac{N_{i,op} + \alpha}{N_{op} + \alpha n}$, where $\alpha$ is the Lidstone smoothing parameter (not to be confused with $\alpha^{\text{max}}$ or $\alpha^{\text{mean}}$ scaling in the tensor features model), with $\alpha=0.25$ in our experiments. $N_{op}$ is the sum of the tf-idf scores of all terms appearing with $op$, $N_{i,op}$ is the sum of the tf-idf scores of all instances of term $i$ appearing with $op$, and $n$ is the number of terms in the vocabulary.

The distribution of operations in the proxy dataset differs from the distribution of operations that appear in \tool\ queries. On GitHub, TensorFlow usage skews toward implementing models and training pipelines, whereas \tool\ queries are tensor manipulations. So, rather than estimating $P(Y_{op})$ from the proxy dataset, we instead use the uniform prior and estimate $P(Y_{op}) = 0.5$ for all operations, which we found to perform better.
The na\"ive Bayes model prioritizes operations with $P(Y_{op} \mid D) > p$, up to $k$ operations, where $p$ and $k$ are hyperparameters.

We experiment with different variations of these models: TF-IDF using $\mathit{minScore} \in \{0.1, 0.15\}$, na\"ive Bayes using $p \in \{0.5, 0.75\}$, and the maximum number of operations prioritized $k\in \{3, 5, 10\}$ for both models. Results for the best settings are shown in Section~\ref{subsec:model_results}.

\section{Experiments}

We now present an evaluation of \tool\ on a set of real-world benchmarks. We use ablation experiments to analyze the overall efficiency gains of \tool's synthesis algorithm compared to baseline approaches. We also perform a study of the synthesis results of \tool\ in comparison to the answers provided on StackOverflow. Finally, we compare \tool's performance with that of human programmers.

\subsection{Benchmark Tasks}
We collected a benchmark set of 70 tensor manipulation tasks, including 50 drawn from StackOverflow questions and 20 real tasks encountered by TensorFlow users in an industrial setting. We aimed to include a diverse set of tasks (sometimes omitting a potential task if it was very similar to an existing benchmark task) as well as a wide range of task difficulties\footnote{Of the 70 benchmark tasks, 9 tasks require the use of 1 operation, 19 tasks use 2 operations, 23 tasks use 3 operations, 11 tasks use 4 operations, 3 tasks use 5 operations, and 5 tasks use 6 or more operations.}. While collecting the benchmark tasks, we noticed that some were not actually amenable to solutions in TensorFlow, so we excluded tasks that we could not solve by hand after much effort. Of the 50 StackOverflow tasks, 34 contained an input/output example in the question. For these problems, we created our own input/output examples inspired by the question poster's example but with numbers changed for licensing reasons. We also expanded these examples (adding more entries to the tensors) where necessary to make the patterns clear. For questions posed without input/output examples, we created examples manually. We also manually wrote single-sentence descriptions for the tasks, striving to use wording that a \tool\ user would plausibly write themselves. We walk through representative instances of our benchmark-creation process below.

\paragraph{User Provides Good Example}
This benchmark comes from the StackOverflow question in Figure~\ref{fig:so_example1}. The user provides an input/output example: the input tensor \code{[45, 58, 72, 33, 45, 58, 58, 33]} should be transformed into the output tensor \code{[0, 1, 2, 3, 0, 1, 1, 3]}. The example has several desirable qualities:
\begin{itemize}
    \item There are no obvious patterns in the choice of numbers in the input tensor. In contrast, if the input tensor were instead \code{[10, 20, 30, 40, 10, 20, 20, 40]}, one could incorrectly construct the output as \code{(in1 / 10) - 1}. In general, we observed that using ``random-looking'' numbers in the input tensor will significantly improve the quality of the example by eliminating coincidental patterns that are not actually relevant to the problem.
    \item There are no obvious patterns in the arrangement of numbers in the input tensor, e.g., the duplicate elements are not all consecutive. This makes it clear that the intended solution must be general enough to handle non-consecutive duplicate elements.
    \item The example tensors have sufficient length. Given only the example, the intended task would be much more ambiguous if the input tensor had, say, 4 elements instead of 8.
    \item The example covers a variety of cases: there are elements appearing exactly 1, 2, and 3 times.
\end{itemize}
Hence, we consider this input/output example to be of high quality.

Even so, due to licensing reasons we change the numbers in the examples for all StackOverflow benchmarks. For this task the input tensor is changed to \code{[37, 42, 42, 37, 28, 15, 42, 15]} and the corresponding output is \code{[0, 1, 1, 0, 2, 3, 1, 3]}, which still satisfies the desirable qualities above. When altering the example tensors, we tried to maintain the ``flavor'' of the example, for instance by preserving the tensor data types and shapes (where the original example was large enough), and by using similar but different numbers (e.g., two digit integers in the input tensor for this problem).

For the natural language description of this task, we use the sentence ``Group items by value and get the group indices.'' Again due to licensing constraints we had to use wording different from that in the original StackOverflow questions, but we tried to use terminology that a real user of \tool\ could plausibly write themselves.

\paragraph{User Provides Ambiguous Example}
This benchmark comes from another StackOverflow question,\footnote{
\url{https://stackoverflow.com/q/51690095}
} where the user wants to gather elements of \code{in2} along axis 1, using indices from \code{in1}. In the user's example \code{in1} is ``\code{[[1], [1]]}'', \code{in2} is ``\code{[[0.2, 0.8], [0.4, 0.6]]}'', and the output is ``\code{[[0.8], [0.6]]}''.

Unfortunately, considering the points from the previous example benchmark, this input/output example is not as good for \tool. The example only includes two ``parts'' (where each part is an element of \code{in2} being indexed), and the same index is used in both parts. Furthermore, the example includes a coincidental pattern -- the extracted elements of \code{in2} are the maximum of each row. Thus, we modify the example and increase the sizes of the tensors to make the intended pattern more clear, while breaking other patterns:

\begin{lstlisting}
in1 = [[2], [0], [1], [0]]
in2 = [[0.2, 0.5, 0.3],
       [0.1, 0.3, 0.6],
       [0.1, 0.6, 0.3],
       [0.7, 0.0, 0.3]]
output = [[0.3], [0.1], [0.6], [0.7]]
\end{lstlisting}

We found that examples posted to StackOverflow are often different in nature from examples created by real \tool\ users. In particular, examples given in StackOverflow questions were often too small because they were intended to be interpreted by humans who also understand the question text, while examples created by actual \tool\ users are much more extensive. This observation is similar to how an example used in a whiteboard discussion with a colleague would be less complete than a battery of examples written for unit tests.

Our single-sentence description of this task is ``Gather elements in a tensor along axis 1.''

\paragraph{User Provides No Example}
In this StackOverflow question,\footnote{
\url{https://stackoverflow.com/q/58466562}
} the user clearly describes the desired behavior but does not provide an input/output example. The user has a tensor \code{in1} of shape $[n, H, W, C]$ containing $n$ images, and a tensor \code{in2} of $n$ scalar weights. The user wishes to multiply each image in \code{in1} by its corresponding scalar weight from \code{in2}.

For such questions without a user-provided input/output example, we create our own example. We make sure that such examples are extensive enough to unambiguously specify the task and simple enough that a \tool\ user could plausibly have written the example. For this task, we use the following:
\begin{lstlisting}
# Shape = [n, H, W, C] = [3, 1, 2, 3].
in1 = [[[[0.1, 0.2, 0.3], [0.4, 0.5, 0.6]]],
       [[[0.8, 1.0, 0.0], [0.6, 0.4, 0.2]]],
       [[[0.9, 0.8, 0.7], [0.1, 0.2, 0.3]]]]
in2 = [2.0, 0.5, 1.0]
output = [[[[0.2, 0.4, 0.6], [0.8, 1.0, 1.2]]],
          [[[0.4, 0.5, 0.0], [0.3, 0.2, 0.1]]],
          [[[0.9, 0.8, 0.7], [0.1, 0.2, 0.3]]]]
\end{lstlisting}

For this task we use the natural language description ``Multiply tensors by scalars in a batched way.''

\subsection{Comparison to \textsc{Transit}}

\begin{figure}[!t]
    \includegraphics[scale=0.5]{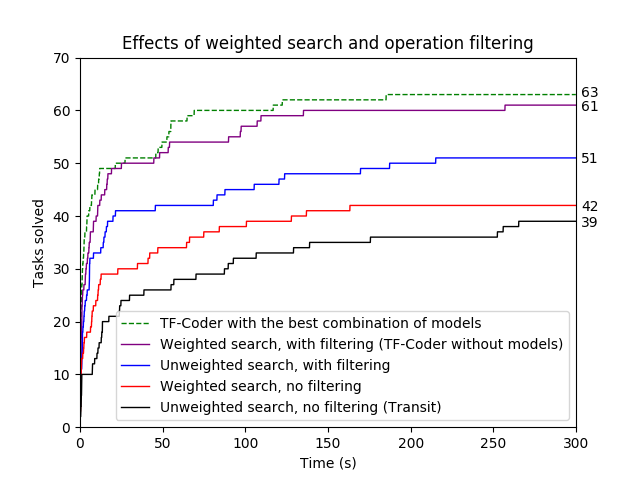}
    \caption{Ablation study investigating the effects of weighted search and operation filtering. The plot shows the number of benchmarks that can be solved within a particular amount of time. Without these improvements, the search algorithm reduces to that of the prior work \textsc{Transit}~\cite{transit}.}
    \label{fig:ablation}
\end{figure}

\tool\ extends the search in \textsc{Transit}~\cite{transit} in several important ways:
\begin{enumerate}
    \item \tool\ incorporates weights for operations and base values, while \textsc{Transit} does not use weights.
    \item \tool\ uses a flexible operation filtering system that generalizes \textsc{Transit}'s type checking, which is insufficient for many TensorFlow operations.
    \item \tool\ uses two models to modify operation weights.
\end{enumerate}
In this section, we evaluate the effectiveness of the first two improvements (the models are evaluated in Section~\ref{subsec:model_results}). We run 4 variations of \tool\ where we independently turn on or off weighting and operation filtering,\footnote{We turn off operation filtering as much as possible, but 37 of 134 operations require filtering to avoid uncatchable segfaults or excessive memory usage.} without using models.

The results of these 4 variations on our benchmarks are plotted in Figure~\ref{fig:ablation}. Both techniques in isolation lead to significant improvement over the \textsc{Transit} algorithm, and their combination produces another large improvement. Overall, \tool\ without any models can solve 61 of the 70 benchmark tasks within 5 minutes, while \textsc{Transit} only solves 39 tasks. Compared to \tool\ without models, removing operation weights results in 10 fewer solved tasks, and removing operation filtering results in 19 fewer solved tasks. It is clear that both of these techniques play a key role in \tool's success.

\paragraph{Perturbing Hardcoded Weights} To see how \tool\ is affected by the choice of hardcoded operation weights, we ran \tool\ where each operation's weight is multiplied by a scaling factor between $0.8$ and $1.25$, drawn from $\exp(\text{Uniform}(\log(4/5), \log(5/4)))$ such that each operation is equally likely to have increased versus decreased weight. Compared to \tool\ with the original weights, the run with perturbed weights solved 1 fewer task, and among the tasks solved by both runs, the perturbed run was 6.2\% slower on average. The perturbed run is only slightly worse, which implies that our weights are chosen reasonably well, but their exact values are not critical to the algorithm's success. Recall that if the weights are removed completely, then \tool\ solves 10 fewer tasks. Thus, we conclude that the existence of reasonable weights are crucial to \tool's success, but their exact values do not need to be chosen perfectly.

\subsection{Effect of the Learned Models}
\label{subsec:model_results}

\begin{table*}
    \caption{The best variations of the tensor features model, natural language model, and combinations, comparing against the baseline of \tool\ without using any models. All methods in this table solved at least the 61 tasks that were solved by the baseline. Models (A), (B), and (C) are tensor features models using cross entropy, $F_1$, and $F_2$ loss functions respectively, each using $\alpha^{\text{max}}$ scaling. Models (X), (Y), and (Z) are natural language models: (X) uses Na\"ive Bayes with $k=3$ and $p=0.5$, (Y) uses Na\"ive Bayes with $k=10$ and $p=0.75$, and (Z) uses TF-IDF with $k=5$ and $\mathit{minScore} = 0.15$. Considering the various metrics, we consider (B) with (Z) to be the best combination of the two model types, marked with $\star$.}
    \setlength{\tabcolsep}{4pt}
	\begin{tabular} {l c c c c c c}
	    \toprule
	           & Tasks & Num faster      & Num slower      & \multicolumn{1}{c}{Time for}    & \multicolumn{1}{c}{Total}   & \multicolumn{1}{c}{Aggregate} \\
	    Model  & solved & (speedup) & (speedup) & \multicolumn{1}{c}{61 tasks (s)} & \multicolumn{1}{c}{speedup} & \multicolumn{1}{c}{speedup} \\ \midrule

	    No models & 61 & --- & --- & 1261.1 & --- & --- \\ \midrule

        \textbf{(A)}: CE, $\alpha^{\text{max}}$ & 62 & 33 (1.88$\times$) & 15 (0.80$\times$) & \phantom{1}979.2 & 1.29$\times$ & 1.38$\times$ \\
        \textbf{(B)}: $F_1$, $\alpha^{\text{max}}$ & \textbf{63} & 38 (1.90$\times$) & 14 (0.79$\times$) & \phantom{-}\textbf{866.1} & \textbf{1.46$\bm\times$} & 1.47$\times$ \\
        \textbf{(C)}: $F_2$, $\alpha^{\text{max}}$ & \textbf{63} & \textbf{41} (1.95$\times$) & \phantom{1}\textbf{7} (0.58$\times$) & 1045.3 & 1.21$\times$ & \textbf{1.54$\bm\times$} \\
        \midrule

        \textbf{(X)}: NB, $k = 3$ & 61 & \textbf{27} (1.69$\times$) & 13 (0.86$\times$) & \textbf{1055.7} & \textbf{1.19$\bm\times$} & 1.20$\times$ \\
        \textbf{(Y)}: NB, $k = 10$ & 61 & 26 (1.73$\times$) & \phantom{1}7 (0.86$\times$) & 1071.9 & 1.18$\times$ & 1.22$\times$ \\
        \textbf{(Z)}: TF-IDF, $k = 5$ & 61 & 23 (1.82$\times$) & \phantom{1}\textbf{7} (0.89$\times$) & 1167.6 & 1.08$\times$ & \textbf{1.27$\bm\times$} \\
        \midrule

        \textbf{(B)} with \textbf{(X)} & \textbf{63} & 43 (2.17$\times$) & 11 (0.77$\times$) & \phantom{1}781.9 & 1.61$\times$ & 1.67$\times$ \\
        \textbf{(B)} with \textbf{(Y)} & \textbf{63} & 42 (2.29$\times$) & 11 (0.78$\times$) & \phantom{-}\textbf{778.0} & \textbf{1.62$\bm\times$} & 1.71$\times$ \\
        \textbf{(B)} with \textbf{(Z)} $\star$ & \textbf{63} & 41 (2.34$\times$) & 11 (0.81$\times$) & \phantom{1}791.2 & 1.59$\times$ & \textbf{1.78$\bm\times$} \\
        \textbf{(C)} with \textbf{(X)} & \textbf{63} & \textbf{47} (2.20$\times$) & \phantom{1}\textbf{5} (0.50$\times$) & \phantom{1}966.1 & 1.31$\times$ & 1.74$\times$ \\
        \bottomrule
	\end{tabular}
    \label{tab:model_results}
\end{table*}

We now evaluate different models to prioritize operations during the enumerative search. We find the best tensor features model (Section~\ref{subsec:tensor_model}) and the best natural language model (Section~\ref{subsec:nl_model}) in isolation, and then find the best combination of the two models under the framework described in Section~\ref{sec:learning}.

To quantify the performance of a synthesizer, a useful metric is the aggregate speedup across multiple tasks compared to a baseline~\citep{speedup}. If a synthesizer solves Task 1 in half the time used by a baseline, and Task 2 in double the time used by the baseline, we view the speedups as ``canceling out.'' Therefore, we compute an aggregate speedup across multiple tasks by first computing the speedup ratio (baseline time) / (synthesizer time) for each task, and then taking the geometric mean of those ratios. This would lead to a speedup ratio of 2$\times$ for the hypothetical Task 1, and 0.5$\times$ for Task 2, which do ``cancel out'' under a geometric mean as desired.

Table~\ref{tab:model_results} lists the performance of the best model variations on our benchmark tasks. For the tensor features models, we experimented with 3 different loss functions and 3 different scaling methods as described in Section~\ref{subsec:tensor_model}. For the natural language models, model-specific hyperparameters are listed as described in Section~\ref{subsec:nl_model}.

Table~\ref{tab:model_results} compares the performance of \tool\ when using each model against \tool\ without any models at all. All model variations and combinations listed in Table~\ref{tab:model_results} solved (at least) all of the 61 benchmark tasks that were solved by the no-model variant. For each task, if using a model results in a solve time that is less than 5\% or less than 0.1 seconds different compared to the no-model run, then we consider the solve times to be ``roughly equal'' and possibly attributed to noise in the timings. The table lists the number of tasks where the timing difference is larger in either direction: ``faster'' means the model does better than not using a model, and ``slower'' means the model does worse.
We also report the aggregate speedup among the faster and slower tasks.
``Time for 61 tasks'' is the sum of the solve times for the tasks that were solved by the no-model variant, and ``total speedup'' compares that total time against that of the no-model variant, as a simple ratio of the two total times. ``Aggregate speedup'' computes the geometric mean of per-task speedup ratios, as described above. Note that ``total speedup'' is biased toward performance on the few difficult long-running tasks, while ``aggregate speedup'' is representative of all tasks (even easy tasks that are solved incredibly quickly where an end-user might not even notice the time saved due to speedups).

Even though \tool\ without models is already highly optimized and powerful, we still observe significant speedups and an increase in the number of solved tasks when using learned models to tweak the search space to better fit the current tensor manipulation task.

\paragraph{Tensor Features Model}
For the tensor features model, we found that $\alpha^{\text{max}}$ scaling was consistently the best scaling approach across all three loss functions. The $F_1$ loss function resulted in the highest total speedup of 1.46$\times$, while the $F_2$ loss function had the highest aggregate speedup of 1.54$\times$. Both of these loss functions solved 2 extra tasks compared to the no-model run.

\paragraph{Natural Language Model}
The best Na\"ive Bayes models obtain higher total speedup than the best TF-IDF model, although the TF-IDF model has slightly better aggregate speedup. Overall, the natural language models were less effective than the tensor features models, but the natural language models lead to fewer and smaller slowdowns and can thus be considered to be more robust.

\paragraph{Model Combinations}
We tried all 9 combinations of the 3 best tensor features models and the 3 best natural language models (as listed in Table~\ref{tab:model_results}), with results for the four best combinations listed at the bottom of the table. The different combinations excel in different ways. Noting the many metrics in Table~\ref{tab:model_results}, as well as performance on the 2 ``extra'' solved tasks, we consider the best combination of models to use $F_1$ loss with $\alpha^{\text{max}}$ scaling as the tensor features model, and TF-IDF with $k = 5$ and $\emph{minScore} = 0.15$ as the natural language model. This combination led to speedups for 41 of 61 tasks (67\%) with an average speedup of 2.34$\times$ for those tasks, and a 1.78$\times$ speedup on average for all tasks. For the 2 additional tasks solved by the use of models, we obtain speedups of 2.5$\times$ and 3.4$\times$ compared to the no-model variant run beyond the 5-minute timeout. Overall, these speedups are a key factor in helping \tool\ feel more interactive.

It is also promising that the model combinations perform significantly better than the individual models alone. This suggests that our framework enables complementary models to jointly influence the enumerative search with compounding benefits.

We also note that Table~\ref{tab:model_results} uses \tool\ without models as a baseline even though it already achieves high performance, solving 61 out of 70 of our benchmark tasks. If we consider that 5 of the unsolved tasks require 6 or more operations which is currently beyond \tool's capability (see Section~\ref{subsec:timeouts}), then there are only 4 unsolved tasks that have reasonably-sized solutions. Adding the models leads to solving 2 of those 4 tasks, which we view as a notable improvement given the context.

A different way of quantifying the effect of models is to add them to a more basic baseline. Because the models prioritize operations by modifying the operations' weights, that baseline must use a weighted search, but operation filtering is not necessary. If we start with the weighted search without filtering and without models (the red line in Figure~\ref{fig:ablation} solving 42 tasks), adding our chosen combination of models leads to solving a superset of 47 tasks, or 5 extra tasks solved, as well as an aggregate speedup of 2.51$\times$ over the 42 tasks solved in both cases.

\subsection{Comparison to StackOverflow Answers}

Since \tool\ was inspired by questions on forums like StackOverflow, it is natural to compare \tool's performance with that of the StackOverflow community. We found that, among the 50 StackOverflow questions, 47 had answers but only 32 had \emph{correct} answers. Incorrect answers included cases where the expert misinterpreted the question, or the solution did not fully generalize, used operations that no longer exist in the current TensorFlow version (TensorFlow 2), or otherwise had bugs that prevent the suggested code from executing successfully. Among correct answers, the median answer-posting time was 31 minutes. In comparison, \tool\ is able to solve 44 of the StackOverflow tasks within 5 minutes, with a median solve time of 1.6 seconds.

\tool's solutions are guaranteed to run successfully on the given example. To examine the generality of \tool's solutions, we manually inspected the solutions and found that they all correctly implement the desired behavior, except for one solution which was mostly correct but had a subtle bug that prevents it from generalizing perfectly. We discuss this in Section~\ref{subsec:buggy}. Even considering this buggy solution, compared to the StackOverflow community, \tool\ improves the success rate from 64\% to 86\% while reducing the median answer time by a factor of over $1000\times$ and reducing the rate of incorrect answers from 32\% to 2\%.

Of course, human-written answers on StackOverflow often contain English explanations for why a solution works, which \tool\ cannot provide. \tool\ is not meant to replace StackOverflow answers in cases where the user needs such an explanation to understand the solution. However, given the comparison above, we believe that \tool\ can still be a very useful tool to TensorFlow users who may be able to utilize \tool's solutions without external explanations, simply by examining the code, running it, and consulting TensorFlow documentation where necessary.

\subsection{Comparison to Human Programmers}
\label{subsec:human}

In this experiment, we evaluated \tool's performance on completely new benchmark tasks, since our choices for the two models was influenced by their aggregate performance on the original set of 70 benchmarks. We examined all StackOverflow questions tagged ``TensorFlow'' in a one-month span, identified those that ask about a tensor manipulation that is in-scope for \tool\ (e.g., \tool\ currently does not support string manipulation or ragged tensors), and filtered out questions that were too easy (answered with a single TensorFlow operation) or too hard (where we were unable to solve the problem with a single TensorFlow expression, despite significant effort). We also excluded problems with boolean outputs, since such questions are difficult to express unambiguously in the form of input/output examples. In the end, we chose 6 new problems intended to be representative of real problems that one might give to \tool. A summary of each problem is provided in Table~\ref{tab:new_problems}. The StackOverflow posts included writing of varying quality, so we wrote our own problem statements to clearly and concisely describe each problem.

\begin{table*}
    \caption{Tasks used in our comparison between \tool\ and human programmers. All times are given in seconds. ``Human Time'' is the median solve time among the 7 human attempts, where an unsuccessful attempt is considered to take more than the time limit of 5 minutes. An example is ``good'' if it causes \tool\ to produce a correct solution within 5 minutes. Among the 7 bad examples, 6 are good after a round of revision (adding missing constants and examining the false positive solutions). ``\tool\ Time'' is the median time it takes \tool\ to produce a correct solution when given the examples after revision.}
    \setlength{\tabcolsep}{4pt}
	\begin{tabular} {c p{5.5cm} | c c | c c c}
	    \toprule
	            &         & Human  & Human & Good     & After    & \tool\ \\
	    \multicolumn{2}{l | }{Problem and Summary} & Solves & Time  & Examples & Revision & Time \\ \midrule
	    A & Concatenate two tensors along axis 1. & \textbf{7/7} & \phantom{> 0}99 & \textbf{3/3} & --- & \phantom{00}\textbf{0.5} \\
	    B & Remove every 3rd element in a vector. & 6/7 & \phantom{> }\textbf{171} & 1/3 & \textbf{3/3} & 230.7 \\
	    C & Gather elements in a batched way. & 0/7 & > 300 & \textbf{3/3} & --- & \phantom{00}\textbf{7.1} \\
	    D & Compute mode along matrix columns. & 1/7 & > 300 & 0/3 & \textbf{3/3} & \phantom{0}\textbf{17.9} \\
	    E & Find a number in a vector and access its index in another vector. & 5/7 & \phantom{> }220 & 1/3 & 2/3 & \phantom{00}\textbf{9.1} \\
	    F & Gather vectors using x and y indices given separately. & 2/7 & > 300 & \textbf{3/3} & --- & \phantom{0}\textbf{33.4} \\
	    \midrule

        \multicolumn{2}{r|}{Overall success rate, median success time} & 50\% & \phantom{> }132 & \textbf{61\%} & 94\% & \textbf{23.3} \\
        \bottomrule
	\end{tabular}
    \label{tab:new_problems}
\end{table*}

Next, we gathered 14 volunteers with TensorFlow experience, including 9 professional programmers and 5 students in machine learning. We asked each volunteer to solve 3 of the 6 problems manually, using whatever tools and resources they would normally use in such a scenario (except \tool), such that each problem was attempted by 7 different people. We imposed a time limit of 5 minutes, not including the time spent reading and understanding the problem statement.

We then asked 3 different people with some familiarity with \tool\ to create an input/output example for each problem, without using \tool\ during that process. The median time to create an example was 87 seconds, again not including the time spent reading the problem statement. Next, we gave the examples to \tool, declaring an example to be ``good'' if it causes \tool\ to produce a correct solution within 5 minutes.

\paragraph{Results} The results for this experiment are shown in Table~\ref{tab:new_problems}. Overall the problems were difficult for the human volunteers to solve---only problem A was solved in all 7 human attempts, problems B and E were solved in the majority of attempts but took several minutes each, and problems C, D, and F were extremely difficult with only 3 successes out of 21 total attempts.

Despite the difficulty of these problems, \tool\ is able to solve all of them when given good examples. In 5 of the 6 problems, the sum of \tool's synthesis time and the median example-creation time of 87 seconds is faster than the median human solve time, although in practice the \tool\ user would also need to spend time understanding \tool's solution and possibly iterating further in the case of false positive solutions.

One question remains: how easy is it to create ``good'' examples for tensor manipulation problems, where the example is robust enough to avoid incorrect \emph{false positive} solutions? This plays a key role in \tool's usability, regardless of how powerful its synthesis capability is. Out of the 18 examples, 11 led to correct solutions, 2 led to timeouts (both on problem B), and 5 led to false positive solutions. In 3 cases (both of the timeouts and one false positive), the example's creator forgot to specify a crucial constant, and adding that constant leads to correct solutions within the time limit. We believe that this kind of mistake is largely eliminated with more experience with \tool. In the remaining 4 false positive cases, we showed \tool's incorrect solution to the example's creator and asked for a revision of the example---3 of the 4 revisions ended up producing correct solutions leading to a final success rate of 94\%. In Section~\ref{subsec:buggy} we examine a representative initial example, the resulting false positive solution, and the successful revision. Even if we only consider the first example, 61\% of the initial examples led to \tool\ producing a correct solution, while only 50\% of human attempts ended in correct solutions.

We also observed that the false positive solutions came exclusively from problems D and E, both of which have the property that the output contains very few elements, all of which appear in the input tensors. Such problems are difficult for programming-by-example systems in general, since there may be many incorrect explanations for the output if it does not contain enough data to rule out those explanations. Even so, \tool\ has a fast turnaround time of under 20 seconds to produce a solution (either a correct solution or a false positive solution) for those problems, allowing the user to quickly return to iterating on the example, and we saw that most of the example revisions led to correct solutions. For the other 4 problems, the example creators consistently produced robust examples to avoid false positive solutions. Hence, we infer that it is not completely trivial to create good examples, but it is still relatively easy for most problems, especially after revising examples when using \tool\ iteratively.

Overall, it is hard to draw confident conclusions (especially with this study's small sample size) in the end-to-end context where \tool\ users must write examples, comprehend \tool's solutions, and possibly iterate in the case of false positive solutions.
Nevertheless, we are impressed by the result that \tool\ solves all 6 problems when given good examples, considering the difficulty of the problems where the median human volunteer could only solve 3. Furthermore, it is promising that 94\% of examples were good after at most one revision. These results suggest that \tool\ is a valuable resource that can help users solve problems that they are unable to solve otherwise. We also hypothesize that \tool\ is best used in a hybrid strategy where the user creates an example, starts a \tool\ search, and in the meantime attempts to solve the problem by hand using the example to help visualize potential solution avenues, execute code snippets, and check handwritten solutions.

\subsection{Comparison to \textsc{AutoPandas}}
\label{subsec:autopandas_comparison}
\textsc{AutoPandas}~\cite{autopandas} is a program synthesizer for the Pandas library, manipulating DataFrames which are similar to tensors. \textsc{AutoPandas} uses a graph representation of DataFrames where edges connect equivalent cells in order to track the flow of data. However, many tensor manipulation problems involve mathematical operations that would break these cell-equivalence edges, so the \textsc{AutoPandas} graph representation is not as well-suited for tensor manipulation problems.

We performed a comparison between \tool\ and \textsc{AutoPandas}, where each tool performs synthesis in the language it was built for (TensorFlow or Pandas), for the same manipulation tasks, to compare the tools' problem-solving abilities. We ran both tools on a machine with one NVIDIA Tesla V100 GPU, but when running \tool, it was faster to disable the GPU and use CPU only. We used a time limit of 20 minutes per task for both tools, as in the \textsc{AutoPandas} paper.

First, we ran \textsc{AutoPandas} on 52 \tool\ benchmarks.\footnote{We could not run \textsc{AutoPandas} on tensors other than rank 2, so we ignored 18 benchmarks involving SparseTensors or tensors of rank 3+ and added dimensions to rank 0 or 1 tensors until they had rank 2.} We turned tensors into DataFrames by using row/column indices as row/column names. \textsc{AutoPandas} solves 11/52 tasks with a mean solve time of 600 seconds. \tool\ solves 45/52 tasks with a mean solve time of 19 seconds. We were surprised that \textsc{AutoPandas} could not solve some relatively easy problems, e.g., the problem in Figure~\ref{fig:problem} can be easily solved in Pandas with \code{in1.div(in1.sum())}, and \textsc{AutoPandas} claims to support both of the operations, but it cannot find a solution. This is evidence that \textsc{AutoPandas}'s graph-based approach is not as effective for tensor manipulation tasks.

Next, we ran \tool\ on \textsc{AutoPandas} benchmarks. Of the 26 tasks, we ignored 10 that we could not solve manually using TensorFlow, mainly those involving SQL-like DataFrame-merging or row-grouping which TensorFlow is not designed to support. Furthermore, DataFrames have row/column names, while tensors do not. The inclusion of these names may make the problem easier (\textsc{AutoPandas} can use names to infer the flow of data) or harder (transforming names into cells and vice versa). We ported the benchmarks to TensorFlow format, thus removing row and column names, while preserving the spirit of the problems as much as we could.\footnote{Some tasks required lambda functions as inputs, and we let \tool\ synthesize that functionality itself. Some I/O examples severely underspecified the intended task (especially after removal of row/column names and lambda functions), so we added entries to the tensors for clarification. We did not use natural language descriptions.}

\tool\ solves 14/16 of the \textsc{AutoPandas} benchmarks with a mean solve time of 56 seconds, while \textsc{AutoPandas} solves 13/16 with a mean time of 149 seconds. \tool\ performs better despite many problems being more difficult to implement in TensorFlow. For instance, one task has a Pandas solution of \code{in1.loc[in2]}, where \code{in1} is a DataFrame and \code{in2} is the function \code{lambda x: x.column\_name != 0}. The TensorFlow solution is more complex, \code{tf.boolean\_mask(in1, tf.cast(in1[:, 1], tf.bool))}, because we must implement the lambda function's behavior in TensorFlow. Even so, \textsc{AutoPandas} times out for this task, while \tool\ solves it in 13 seconds.

When comparing \tool\ and \textsc{AutoPandas}, an important piece of context is the relative size of the search spaces considered by \tool\ and \textsc{AutoPandas}. It is difficult to directly compute or compare these search space sizes, but as a reference point, \tool\ supports 134 operations while \textsc{AutoPandas} supports 119 operations. Furthermore, all \textsc{AutoPandas} benchmarks can be solved using at most 3 Pandas operations, and the Pandas solutions use 1.9 operations on average over the 16 \textsc{AutoPandas} benchmarks we considered. In contrast, the simplest TensorFlow solutions to our knowledge (using operations supported by \tool) use on average 3.4 operations for those benchmarks. Hence, it is likely that \tool\ must consider a larger search space with more available operations and more complex programs compared to \textsc{AutoPandas}, when running on the \textsc{AutoPandas} benchmarks. This search space comparison makes \tool's superior performance more impressive.

In summary, \tool\ outperforms \textsc{AutoPandas} by solving more tasks with faster solve times on both sets of benchmarks.

\subsection{Public Tool}
We have open-sourced \tool\ and made it available online for public use.\footnote{\url{https://github.com/google-research/tensorflow-coder}} Users may opt in to have their usage data recorded; we do not collect any data for users who do not opt in. At the time of writing, among the recorded data, \tool\ successfully found solutions to 415 unique problems (73\% of which are solved in under 10 seconds), while 194 unique problems were unsolved. However, not all of the unsolved problems indicate poor synthesizer performance. We manually inspected a random selection of 50 of the 194 unsolved problems, finding the following:
\begin{itemize}
    \item \emph{14 successes after fixing settings.} These problems can actually be solved by changing the online tool's default time limit of 1 minute to 5 minutes as in our experiments, and/or changing the tool setting indicating whether \tool\ is required to use every given input in the solution (users sometimes provide extraneous inputs but choose the setting where all inputs are required to appear in the solution).
    \item \emph{14 bad examples.} These problems contain a bad I/O example, either with a numeric error in the output tensor that deviates from the intended pattern, or with necessary input tensors or constants missing from the specification. Some of these problems appear to come from users ``testing'' the tool with intentionally incorrect examples modified from correct ones.
    \item \emph{1 unclear problem.} This problem has an I/O example where we cannot find a reasonable pattern, and the natural language description of the task is empty. Even though we do not know what the intended transformation is, \tool\ does find a solution within 5 minutes, but we speculate that this solution is a false positive because it does not make intuitive sense.
    \item \emph{21 true failures.} For these problems, we manually confirmed that a solution exists within \tool's search space but \tool\ is unable to find a solution within 5 minutes. Many of these problems are very difficult; 13 of the 21 problems use 6 or more operations in the simplest handwritten solutions to our knowledge, but the largest \tool\ solution among our benchmarks contains 5 operations.
\end{itemize}
If we treat the first category ``fixing settings'' as successes, ignore the ``bad examples,'' and treat the ``unclear problem'' as a ``true failure,'' we can extrapolate from the 50 randomly-selected problems and estimate that the 194 unsolved problems contain approximately 54 successes and 85 true failures. Combining these with the 415 problems that \tool\ did solve, we estimate that \tool\ obtains a success rate of about 85\% on well-specified problems ``in the wild,'' which is only slightly lower than \tool's 90\% success rate on our 70 benchmark tasks. We are encouraged by \tool's ability to solve many unique user-generated problems and its high success rate in public usage.

\section{Analysis of Synthesized Programs}

In this section we take a closer look at the programs synthesized by \tool\ to analyze its strengths and weaknesses.

\subsection{A Selection of Synthesized Programs}

\begin{figure}
    \centering
    \begin{subfigure}[b]{\linewidth}
        \begin{lstlisting}
# Convert tensor into pairs for SparseTensor indexing.
in1 = [0, 0, 0, 1, 3, 3]
output = [[0, 0], [0, 1], [0, 2], [1, 0], [3, 0], [3, 1]]

# Solution found in 2.6 seconds
tf.cast(tf.where(tf.sequence_mask(tf.math.bincount(in1))), tf.int32)
\end{lstlisting}
        \vskip -8pt
        \caption{A real task from an industrial setting that is incredibly tricky to solve, using an unintuitive composition of uncommon operations.}
        \label{subfig:google_01}
    \end{subfigure}

    \vskip 4pt
    \begin{subfigure}[b]{\linewidth}
        \begin{lstlisting}
# Reorder segments.
in1 = [10, 20, 30, 40, 50, 13, 17, 19, 21, 22, 23]
in2 = [1, 1, 1, 1, 1, 0, 0, 0, 2, 2, 2],
output = [13, 17, 19, 10, 20, 30, 40, 50, 21, 22, 23]

# Solution found in 2.5 seconds
tf.gather(in1, tf.argsort(in2, axis=0, stable=True))
\end{lstlisting}
        \vskip -8pt
        \caption{Another task from an industrial setting. The use of \code{tf.argsort} is crucial---the problem would be very difficult without it. \tool\ can help users learn about operations that they are unfamiliar with.}
        \label{subfig:google_09}
    \end{subfigure}

    \vskip 4pt
    \begin{subfigure}[b]{\linewidth}
        \begin{lstlisting}
# Linear interpolation between two tensors.
in1 = [[[1.0, 2.0], [3.0, 4.0], [5.0, 6.0]],
        [[10., 20.], [30., 40.], [50., 60.]]],
in2 = [[[9.0, 8.0], [7.0, 6.0], [5.0, 4.0]],
       [[90., 80.], [70., 60.], [50., 40.]]]
in3 = [0.1, 0.4, 0.8]
output = [[[8.2, 7.4], [5.4, 5.2], [5.0, 5.6]],
          [[82., 74.], [54., 52.], [50., 56.]]]

# Solution found in 123.2 seconds
tf.add(in2, tf.multiply(tf.expand_dims(in3, 1), tf.subtract(in1, in2)))
\end{lstlisting}
        \vskip -8pt
        \caption{A StackOverflow task with three inputs. Our best handwritten solution used 6 operations, while \tool's solution only uses 4. \tool's solutions are often more elegant than those created by experienced TensorFlow programmers.}
        \label{subfig:stackoverflow_35}
    \end{subfigure}

    \vskip 4pt
    \begin{subfigure}[b]{\linewidth}
        \begin{lstlisting}
# Convert segment lengths to segment ids.
in1 = [3, 4, 1]
output = [0, 0, 0, 1, 1, 1, 1, 2]

# Solution found in 4.0 seconds
tf.cast(tf.where(tf.sequence_mask(in1))[:, 0], tf.int32)
\end{lstlisting}
        \vskip -8pt
        \caption{A StackOverflow task that is surprisingly difficult in TensorFlow. The answer posted to StackOverflow uses 9 operations while \tool\ only needs 4, again demonstrating \tool's ability to find simple solutions.}
        \label{subfig:stackoverflow_46}
    \end{subfigure}

    \vskip 4pt
    \begin{subfigure}[b]{\linewidth}
        \begin{lstlisting}
# Find the indices of all elements.
in1 = [32, 53, 45, 38, 29, 89, 64, 23]
in2 = [38, 53, 89, 38, 32, 64]
output = [3, 1, 5, 3, 0, 6]

# Solution found in 65.6 seconds
tf.cast(tf.argmax(tf.cast(tf.equal(in1, tf.expand_dims(in2, 1)), tf.int32), axis=1), tf.int32)
\end{lstlisting}
        \vskip -8pt
        \caption{This StackOverflow task requires a particularly long solution, involving 5 TensorFlow operations and 11 nodes in the expression tree. It is impressive that \tool\ finds solutions of this size, considering \tool's enormous search space with 134 supported operations.}
        \label{subfig:stackoverflow_48}
    \end{subfigure}

    \vskip -4pt
    \caption{Results on selected tasks (descriptions in comments). None of these tasks used user-provided constants.}
    \label{fig:selected_results}
    \vskip -6pt
\end{figure}

Figure~\ref{fig:selected_results} shows examples of interesting problems that \tool\ is able to solve. In the figure, problems (a) and (b) come from the industrial setting, while problems (c) through (e) are drawn from StackOverflow questions.

For problems (a), (b), (c), (d), and several other problems in our benchmark suite, \tool\ identifies a solution that is more elegant than solutions created by experienced TensorFlow programmers. In particular, problem (b) was a real task encountered by an author of this paper (while working on a different project), and their initial attempt at solving the problem resulted in a 12-line function using 18 TensorFlow operations. The problem was discussed with a colleague, who then proposed a different solution in the form of another 12-line function using 17 TensorFlow operations. Although the problem is simple to explain, it is evidently extremely difficult for humans to solve, even for professional deep learning researchers.

Incredibly, \tool\ solves this problem in 2.5 seconds using only 2 operations. It turns out that \code{tf.argsort(values, stable=True)} is especially helpful for this problem, although it is not obvious that this is the case. Even if a human can solve a problem themselves, having a simpler solution makes their code faster, more understandable, and easier to maintain, as was actually the case for problem (b).

Due to the exhaustive nature of \tool, it can find elegant solutions to problems using uncommon operations that a human programmer might not know about, or unconventional combinations of operations that the programmer might not have considered. Problem (a) is another example of this, where \tool's solution uses the result of \code{tf.math.bincount} as the argument to \code{tf.sequence\_mask}. This composition of operations is unusual, and in fact there are zero instances of the code ``\code{tf.sequence\_mask(tf.math.bincount}'' in publicly searchable GitHub repositories, in StackOverflow questions or answers, or in Google search results.

Considering unconventional combinations of operations is a desirable behavior of \tool\ because it leads to simple and elegant solutions that humans would otherwise overlook. However, this behavior would not be expected from other synthesis approaches that attempt to imitate existing code corpora (e.g., a generative approach using a neural sequence model), since such combinations of operations might not appear at all in the dataset.

Problem (e) shows one of the largest solutions found by \tool, involving 5 TensorFlow operations and 11 total nodes in the expression tree. In \tool's search space for this problem (which includes 134 different operations, 2 input tensors, and 11 different constants), there are $2.3 \times 10^{18}$ expressions using at most 5 operations and 11 nodes. Despite this colossal search space, \tool\ is able to find this solution in only one minute.

\subsection{False Positive Solutions}
\label{subsec:buggy}

As with any programming-by-example (PBE) system, \tool\ sometimes produces \emph{false positive} solutions, i.e., solutions that work for the given example but fail to generalize in the intended way. It is possible for a false positive solution to look reasonable enough that a user mistakes it for a correct solution, but in our experience, false positive solutions are most often noticeably wrong. In any case, we stress that users should carefully review solutions produced by \tool\ before using them in real projects, as one should do for any code produced by a PBE system or other coding tools. In this section we discuss two instances of false positives produced by \tool.

\paragraph{On the 70 benchmark tasks}
When running \tool\ on our 70 benchmark tasks, 62 of its 63 solutions are correct. We now examine the single false positive solution.

In this task from the industrial setting, the user wants to sum elements of \code{in1}, but partitioned into groups specified by \code{in2} first. The user provides the following example, which we use as-is in our benchmark task without modification:
\begin{lstlisting}
in1 = [5, 7, -12, 10, 20]
in2 = [1, 2, 3, 1, 2]
output = [15, 27, -12, 15, 27]
\end{lstlisting}
In this example, the elements \code{5} and \code{10} of \code{in1} are both in group~\code{1} (specified by \code{in2}), so their sum, \code{15}, is present in the corresponding positions in the output. Considering the format of \code{in2} as provided by the user, we assume that it will only contain integers from 1 to $G$ inclusive, if there are $G$ distinct groups.

\tool's solution to this problem is:
\begin{lstlisting}
tf.gather(tf.math.unsorted_segment_sum(in1, in2, tf.reduce_sum(in1)), in2)
\end{lstlisting}
This is very close to being a correct solution, but it does have a subtle bug. The TensorFlow operation \code{tf.math.unsorted\_segment\_sum(data, segment\_ids, num\_segments)} is useful here, taking care of grouping and summing, but it requires that \code{num\_segments} be sufficiently large (but being too large will hinder efficiency). For this particular I/O example, setting \code{num\_segments} to \code{tf.reduce\_sum(in1)} happens to be large enough so the solution works in this case, but this is not true in general (e.g., if \code{in1} were entirely negative). A bug-free solution would use \code{tf.reduce\_max(in2) + 1} instead:
\begin{lstlisting}
tf.gather(tf.math.unsorted_segment_sum(in1, in2, tf.reduce_max(in2) + 1), in2)
                                                 ^^^^^^^^^^^^^^^^^^^^^^
\end{lstlisting}

Although \tool's solution was not perfect, it was nearly so, such that a human user reviewing the solution (while looking at TensorFlow documentation if needed) could identify the bug and write a fix. Hence, we view \tool\ as still being helpful for this scenario since it identifies a clear path toward a solution.

\paragraph{On the 6 additional tasks}
In our comparison between \tool\ and human programmers in Section~\ref{subsec:human}, we gathered 6 more tasks beyond the original 70 benchmark problems. We asked 3 \tool\ users to create examples for these 6 tasks, but 4 of the 18 input/output examples were not unambiguous enough and led to false positive solutions. The following is a discussion of a representative false positive scenario.

The task is to compute the mode of a 2D tensor along its columns. The initial input/output example was:
\begin{lstlisting}
in1 = [[1, 3, 2, 9, 5],
       [1, 7, 2, 4, 6],
       [2, 3, 4, 9, 6],
       [1, 7, 4, 9, 6],
       [1, 5, 2, 9, 6],
       [1, 7, 2, 5, 4]]
output = [1, 7, 2, 9, 6]
\end{lstlisting}
From this example, \tool\ produces the following false positive solution:
\begin{lstlisting}
tf.reduce_max(tf.gather(in1, (0, 1)), axis=0)
\end{lstlisting}
It is quite obvious that this solution is not correct---it simply takes the maximum in each column among the first two rows of \code{in1}. After seeing this false positive solution, the user revises the example by adding a new column where the mode is not the maximum among the first two rows:
\begin{lstlisting}
in1 = [[1, 3, 2, 9, 5, 8],
       [1, 2, 2, 4, 6, 5],
       [2, 3, 4, 9, 6, 5],
       [1, 2, 4, 9, 6, 7],
       [1, 5, 2, 9, 6, 3],
       [1, 2, 2, 5, 4, 5]]
output = [1, 2, 2, 9, 6, 5]
\end{lstlisting}
The added column rules out the false positive, and \tool\ finds a correct solution instead\footnote{The problem assumes that the elements are labels for 10 categories, represented by integers from 0 to 9. The number of categories (10) is allowed as a constant.}:
\begin{lstlisting}
tf.cast(tf.argmax(tf.reduce_sum(tf.one_hot(in1, 10), axis=0), axis=1), tf.int32)
\end{lstlisting}
This scenario illustrates how \tool\ can be used iteratively. The first solution is easily recognized as a false positive, and it is then straightforward to revise the example to avoid that false positive.

This problem is actually somewhat prone to false positives because the output has one element selected from each column of the input, and each selected element appears many times in the corresponding input column. In the original example, if one number is randomly selected from each column, there is a 12\% chance of producing the correct output. There are many possible procedures for choosing an element from each column, and it is quite possible that one such procedure (e.g., ``take the maximum among the first 2 rows'') chooses the correct numbers using the wrong logic, leading to a false positive. With this understanding, one can strengthen the example even further by including columns where the mode only appears twice and every other number in the column appears once. We recognize that creating robust examples is not trivial, but it does become easier with experience after recognizing common sources of false positives and how to mitigate them, such as the scenario described here.

\subsection{Timeouts}
\label{subsec:timeouts}
Among the 70 benchmark tasks, 7 tasks cause \tool\ to time out after 5 minutes. All of the 7 unsolved problems are quite difficult, requiring 3, 5, 6, 9, 10, 10, and 13 operations in the simplest solutions to our knowledge. The easiest of these problems (3 operations, 2 of which are quite uncommon) is solved correctly in 31 minutes, even though none of the required operations were prioritized by the models. We believe a solution could be found much faster if the models predicted correctly for this task. The 5-operation problem is solved correctly in 58 minutes, and 3 of the required operations were prioritized by the tensor features model. Interestingly, both of these solutions required one fewer operation than our best handwritten solutions, showing again that \tool\ can find simpler solutions compared to professional TensorFlow users. The remaining problems are simply too difficult for \tool\ to solve within an hour considering the exponentially-sized search space with respect to the solution length.

\section{Related Work}
In this section, we discuss related works on programming by examples and program synthesis from several domains.

\paragraph{Programming By Example (PBE)} The problem of synthesizing programs from input/output examples has been studied for a long time starting with the approaches of synthesizing LISP programs~\cite{shaw, hardy}. More recently, PBE techniques have been developed for domains including string transformations~\cite{flashfill, cacm12, blinkfill}, data extraction from semi-structured formats~\cite{flashextract}, data structure manipulations~\cite{lambdasquare, storyboard}, distributed cache coherence protocols~\cite{transit},
data imputation programs~\cite{WangDS17,WangDS18}, map-reduce programs~\cite{SmithA16}, and Java functions~\cite{frangel}.

Unlike these approaches, which synthesize programs from only input/output examples, \tool\ uses both input/output examples and natural language descriptions to guide a weighted enumerative search. \citet{SketchRegex} present a technique to generate regular expressions from natural language and examples, where the natural language is first used to generate a program sketch and the sketch is then completed using an enumerative approach using examples. On the other hand, \tool\ uses both examples and natural language simultaneously to guide a weighted bottom-up search over compositions of supported operations. \textsc{Synquid}~\cite{synquid} also uses type-based reasoning and filtering for synthesis. In contrast, \tool\ uses dynamic value-based checks for argument and combination filters for different TensorFlow operations.

\paragraph{Machine Learning for Program Synthesis}
With the recent advances in machine learning, there has been much interest in using such techniques for program synthesis. RobustFill~\cite{robustfill,nsps} uses an encoder-decoder model to generate string transformation programs from examples. The encoder embeds the example strings using recurrent LSTM networks, which is then used to condition the output program sequence decoder. DeepCoder~\cite{deepcoder} trains a model to learn a distribution over possible list functions given the input/output list examples. It then uses the distribution to guide an enumerative search. \textsc{Euphony}~\cite{euphony} performs a weighted enumerative search using the A* search algorithm, where the weights come from a probabilistic higher-order grammar (PHOG). 

Similar to these previous approaches, \tool\ also learns a distribution over possible programs conditioned on the corresponding specification and then uses it for efficiently searching the desired program. However, there are some key differences that can be described along several dimensions. First, these approaches only use input/output examples as the specification to learn the program distribution whereas \tool\ uses both input/output examples and natural language descriptions for specification. Second, while RobustFill and \textsc{Euphony} learn weights/probabilities of program expansion conditioned on both the specification and the partial program (generated during the search), DeepCoder and \tool\ learn weights conditioned on only the specification. Third, these approaches also incorporate weights differently into program search. RobustFill and \textsc{Euphony} perform a weighted top-down search/program expansion, while DeepCoder supports different search options such as top-down DFS (depth-first search), sort-and-add (DFS with only active functions), and constraint solving using the \textsc{Sketch} synthesizer. In contrast to these search mechanisms, \tool\ performs a weighted bottom-up search where each partial program is executable, which also enables efficient equivalence-based pruning.
Finally, \textsc{Euphony} uses non-neural techniques to learn code patterns from given solutions to easy problems which then facilitate solving harder problems, whereas DeepCoder, RobustFill, and \tool\ all use neural networks to tackle individual problems without requiring a set of known solutions to similar problems.

\textsc{AutoPandas}~\cite{autopandas} uses graph neural networks to synthesize Pandas programs that manipulate DataFrames, which are similar to TensorFlow tensors. A key innovation in \textsc{AutoPandas} is a graph representation of the input and output DataFrames with edges connecting equal cells. Although tensors and DataFrames are similar, \textsc{AutoPandas}' graph approach is not as applicable to the TensorFlow domain, since many common mathematical operations would break the cell-equivalence edges. In other words, DataFrames retain much of their data while being manipulated through pivots, selections, and joins, making it easy for cell-equivalence edges to track the movement of data, while this is only true for a fraction of manipulations in TensorFlow. In Section~\ref{subsec:autopandas_comparison} we found that \tool\ outperforms \textsc{AutoPandas} on both sets of benchmarks, from \tool\ and from \textsc{AutoPandas}.

There are also some approaches that use machine learning for ranking programs. FlashFill uses version-space algebra to identify all programs in a DSL that are consistent with a set of input/output examples, and then uses a ranking function learned through supervised learning~\cite{rankpbe} to rank the programs, so that the user does not need to provide too many examples before obtaining the desired program. Unlike this ranking approach that first finds all consistent programs, \tool\ uses learning to guide the search in first place.

\citet{menon} describe an approach for synthesizing string manipulation programs that learns a probabilistic context free grammar (PCFG) of rules given a set of examples. It uses a set of hand-designed clues to learn a distribution over likely rules and then enumerates over a subset of rules in order of decreasing probabilities to search for a consistent program. Since it learns from a small number of training examples (280), the clues need to be very domain-specific. In comparison, \tool's TensorFlow domain is quite different from the string-processing domain. \tool\ trains a model to learn a distribution over operations from millions of synthetically generated programs, and the model is used to guide an efficient weighted enumerative search with value- and type-based filtering and pruning strategies.

\paragraph{Automated Machine Learning} AutoML techniques~\cite{zoph2016neural,thornton2013auto} use learning and evolutionary approaches to automatically generate machine learning model architectures that achieve the best accuracy on a training dataset. The key idea in these approaches is to parameterize the space of model architectures in terms of different choices for the number of neural network layers, hidden units, activation functions, loss function, etc., and then efficiently search this space to discover architectures that lead to better validation set accuracy. AL~\cite{al} learns to generate end-to-end supervised learning pipelines from a training corpus of 500 pipelines and their corresponding datasets. The key idea in AL is that for many machine learning tasks, the data featurization and modeling choices could be transferred and reused across different pipelines. In contrast to the AutoML approaches that aim to automatically search for the best-performing model architectures, \tool\ instead uses input/output examples to synthesize implementations of tensor manipulations that machine learning practitioners perform while designing complex deep learning computations.

\paragraph{Program Synthesis}
There has been a renewed interest in program synthesis research in the last decade because of the advances in both constraint solving and algorithmic synthesis techniques~\cite{sygus,synthesissurvey}. The synthesis approaches can be broadly classified based on the underlying search mechanism: (i) enumerative~\cite{transit}, (ii) constraint-based~\cite{sketch1,sketch2}, and (iii) stochastic~\cite{stoke, frangel}. Applying constraint-based synthesis techniques to the TensorFlow domain would require a huge effort of modeling semantics of TensorFlow operations, and for many operations these would not be scalable due to complex non-linear computations. \tool\ builds on top of the bottom-up enumerative search from \textsc{Transit}~\cite{transit}, adding expression weights and flexible value-based filtering for a more efficient search. Moreover, it dynamically adjusts weights using learned models based on the input/output examples and natural language description.

\textsc{Probe}~\cite{probe} also extends the \textsc{Transit} algorithm, performing a bottom-up search guided by learning from programs that satisfy a subset of examples. The two key ideas in \textsc{Probe} are adding weights to a bottom-up search (an idea also in \tool\ but developed concurrently\footnote{\tool\ and \textsc{Probe} are concurrent work, with \tool\ appearing on arXiv 7 months earlier than \textsc{Probe}.}), and learning such weights on-the-fly from ``promising partial solutions'' which are programs that pass a subset of the examples. The \textsc{Probe} technique as described in the paper uses multiple input/output examples of varying difficulty, such that one can solve a subset of the examples with shorter code than the full solution. Even though \tool\ only uses a single example, due to the composite nature of tensors we can often use different parts of the example to act as multiple smaller examples.

However, we note that solutions to subsets of examples can only help the synthesizer find the full solution if the problem has the ``special-case similarity'' property~\cite{frangel}, which asserts the existence of special cases (subsets of examples) that can be solved by \emph{similar and simpler} code compared to the full solution. Some tensor manipulation programs have the special-case similarity property, for example the task in Figure~\ref{fig:problem} where the \code{tf.reduce\_sum(in1, axis=1)} portion can be simplified to a constant and still be correct for rows that sum to that constant. However, we observe that such special-case similarity is not as common for the tensor manipulation domain. One reason for this is that the huge variety of operations allows many tasks to be solved by composing few operations. In other words, the search space is extremely broad but is not searched to as much depth, often resulting in simple-but-clever solutions which reduces the opportunities for partially-correct solutions to be even simpler. Another reason is that special-case similarity occurs more often in domains such as string manipulation where many operations could be no-ops in the right scenarios. For example, string concatenation is a no-op when one of the arguments is the empty string. This property leads to simpler solutions omitting that operation for the subset of examples producing these scenarios. But because most tensor-manipulating operations change the tensor's shape or data type, it is much less common for those operations to possibly be no-ops. Since special-case similarity is not as common for tensor manipulation as it is for some other domains, we expect that the effectiveness of learning from partially-successful solutions would be reduced in this domain.

\section{Conclusion}
In this paper, we presented \tool, a synthesis tool for automatically generating tensor manipulation programs in TensorFlow from examples and natural language. \tool\ employs a bottom-up weighted enumerative search with type- and value-based filtering to conform to the constraints imposed by TensorFlow operations. It uses two machine learning models to predict useful operations from features of the input/output tensors and a natural language description of the task, and these predictions are combined within a general framework to modify the weights to customize the search process for the given task. We evaluated \tool\ on several real-world tensor transformation tasks faced by TensorFlow users on StackOverflow and in an industrial setting, and various ablation experiments highlight the usefulness of the two models and filtering techniques. We also found that \tool\ outperforms the prior work \textsc{AutoPandas}, it finds solutions to StackOverflow questions that human TensorFlow programmers struggle to solve, and it sometimes finds solutions that are simpler than the best handwritten solutions written by professional deep learning researchers. We believe that \tool\ can help both machine learning beginners and experienced practitioners in writing tricky tensor transformation programs that are common in deep learning pipelines.

Perhaps the most important lesson to be learned from this work is simply the fact that a well-optimized enumerative search can successfully solve real-world tensor manipulation problems within seconds, even on problems that human programmers struggle to solve within minutes.

\begin{acks}
  The authors thank Charles Sutton and the other members of the program synthesis team at Google Brain for helpful discussions.
\end{acks}

\bibliographystyle{ACM-Reference-Format}
\bibliography{tfcoder}

\clearpage

\appendix

\section{Supported Operations in \tool}
\label{app:ops}
Below is the list of 134 operations currently supported by \tool. We did not cherrypick the operations to support; in fact, out of the 134 supported operations, only 59 are used in \tool's solutions to our benchmark tasks.

\begin{lstlisting}[keywordstyle=\ttfamily\footnotesize,basicstyle=\linespread{0.9}\ttfamily\footnotesize]
General TensorFlow functions:
-----------------------------
tf.abs(x)
tf.add(x, y)
tf.add_n(inputs)
tf.argmax(input, axis)
tf.argmin(input, axis)
tf.argsort(values, axis, stable=True)
tf.argsort(values, axis, direction='DESCENDING', stable=True)
tf.boolean_mask(tensor, mask)
tf.broadcast_to(input, shape)
tf.cast(x, dtype)
tf.clip_by_value(t, clip_value_min, clip_value_max)
tf.concat(values, axis)
tf.constant(value)
tf.constant(value, dtype)
tf.divide(x, y)
tf.equal(x, y)
tf.exp(x)
tf.expand_dims(input, axis)
tf.eye(num_rows)
tf.eye(num_rows, num_columns)
tf.eye(num_rows, dtype)
tf.fill(dims, value)
tf.gather(params, indices)
tf.gather(params, indices, axis, batch_dims)
tf.gather_nd(params, indices)
tf.gather_nd(params, indices, batch_dims)
tf.greater(x, y)
tf.greater_equal(x, y)
tf.math.bincount(arr)
tf.math.ceil(x)
tf.math.count_nonzero(input)
tf.math.count_nonzero(input, axis)
tf.math.cumsum(x, axis)
tf.math.cumsum(x, axis, exclusive=True)
tf.math.divide_no_nan(x, y)
tf.math.floor(x)
tf.math.log(x)
tf.math.negative(x)
tf.math.reciprocal(x)
tf.math.reciprocal_no_nan(x)
tf.math.segment_max(data, segment_ids)
tf.math.segment_mean(data, segment_ids)
tf.math.segment_min(data, segment_ids)
tf.math.segment_prod(data, segment_ids)
tf.math.segment_sum(data, segment_ids)
tf.math.squared_difference(x, y)
tf.math.top_k(input, k)
tf.math.unsorted_segment_max(data, segment_ids, num_segments)
tf.math.unsorted_segment_mean(data, segment_ids, num_segments)
tf.math.unsorted_segment_min(data, segment_ids, num_segments)
tf.math.unsorted_segment_prod(data, segment_ids, num_segments)
tf.math.unsorted_segment_sum(data, segment_ids, num_segments)
tf.matmul(a, b)
tf.maximum(x, y)
tf.minimum(x, y)
tf.multiply(x, y)
tf.not_equal(x, y)
tf.one_hot(indices, depth)
tf.ones(shape)
tf.ones_like(input)
tf.pad(tensor, paddings, mode='CONSTANT')
tf.pad(tensor, paddings, mode='CONSTANT', constant_values)
tf.pad(tensor, paddings, mode='REFLECT')
tf.pad(tensor, paddings, mode='SYMMETRIC')
tf.range(start)
tf.range(start, limit, delta)
tf.reduce_any(input_tensor, axis)
tf.reduce_max(input_tensor)
tf.reduce_max(input_tensor, axis)
tf.reduce_mean(input_tensor)
tf.reduce_mean(input_tensor, axis)
tf.reduce_min(input_tensor)
tf.reduce_min(input_tensor, axis)
tf.reduce_prod(input_tensor, axis)
tf.reduce_sum(input_tensor)
tf.reduce_sum(input_tensor, axis)
tf.reshape(tensor, shape)
tf.reverse(tensor, axis)
tf.roll(input, shift, axis)
tf.round(x)
tf.searchsorted(sorted_sequence, values, side='left')
tf.searchsorted(sorted_sequence, values, side='right')
tf.sequence_mask(lengths)
tf.sequence_mask(lengths, maxlen)
tf.shape(input)
tf.sign(x)
tf.sort(values, axis)
tf.sort(values, axis, direction='DESCENDING')
tf.sqrt(x)
tf.square(x)
tf.squeeze(input)
tf.squeeze(input, axis)
tf.stack(values, axis)
tf.subtract(x, y)
tf.tensordot(a, b, axes)
tf.tile(input, multiples)
tf.transpose(a)
tf.transpose(a, perm)
tf.unique_with_counts(x)
tf.unstack(value, axis)
tf.where(condition)
tf.where(condition, x, y)
tf.zeros(shape)
tf.zeros_like(input)


SparseTensor functions:
-----------------------
tf.SparseTensor(indices, values, dense_shape)
tf.sparse.add(a, b)
tf.sparse.concat(axis, sp_inputs)
tf.sparse.expand_dims(sp_input, axis)
tf.sparse.from_dense(tensor)
tf.sparse.maximum(sp_a, sp_b)
tf.sparse.minimum(sp_a, sp_b)
tf.sparse.reduce_max(sp_input, axis, output_is_sparse)
tf.sparse.reduce_sum(sp_input, axis, output_is_sparse)
tf.sparse.reset_shape(sp_input)
tf.sparse.reshape(sp_input, shape)
tf.sparse.retain(sp_input, to_retain)
tf.sparse.slice(sp_input, start, size)
tf.sparse.split(sp_input, num_split, axis)
tf.sparse.to_dense(sp_input)
tf.sparse.to_dense(sp_input, default_value)
tf.sparse.to_indicator(sp_input, vocab_size)
tf.sparse.transpose(sp_input)
tf.sparse.transpose(sp_input, perm)


Python-syntax operations:
-------------------------
IndexingAxis1Operation:             arg1[:, arg2]
IndexingOperation:                  arg1[arg2]
PairCreationOperation:              (arg1, arg2)
SingletonTupleCreationOperation:    (arg1,)
SlicingAxis0BothOperation:          arg1[arg2:arg3]
SlicingAxis0LeftOperation:          arg1[arg2:]
SlicingAxis0RightOperation:         arg1[:arg2]
SlicingAxis1BothOperation:          arg1[:, arg2:arg3]
SlicingAxis1LeftOperation:          arg1[:, arg2:]
SlicingAxis1RightOperation:         arg1[:, :arg2]
TripleCreationOperation:            (arg1, arg2, arg3)
\end{lstlisting}

\end{document}